\documentclass[11pt]{spie}  
\usepackage[]{graphicx}

\usepackage{dsfont,amsmath,amsfonts,subfig,color}
\def\argmin{\mathop{\rm argmin}}
\usepackage{hyperref}
\usepackage{cleveref}
\Crefformat{figure}{#2Fig.~#1#3}
\Crefformat{table}{#2Tab.~#1#3}
\Crefmultiformat{figure}{Figs.~#2#1#3}{ and~#2#1#3}{, #2#1#3}{ and~#2#1#3}

\begin{document} 

\centerline{\Large \bf
Novel Sparse Recovery Algorithms for 3D Debris Localization}
\vspace{5pt}
  \centerline{\Large \bf  using}
\vspace{5pt}
\centerline{\Large \bf  Rotating Point Spread Function Imagery\footnote[1]{Dedicated to the memory of Professor Mila Nikolova who passed away on June 20th, 2018. }}
\vspace{30pt}

\centerline{\large \bf Chao Wang}
\centerline{\large \it Dept. Mathematics, Chinese Univ. of Hong Kong, Shatin, Hong Kong}

\vspace{8pt}
\centerline{\large \bf Robert Plemmons}
\centerline{\large \it Dept. Computer Science, Wake Forest Univ., Winston-Salem, NC 27109}

\vspace{8pt}
\centerline{\large \bf Sudhakar Prasad\footnote[2]{Now at School of Physics and Astronomy, Univ. Minnesota, Minneapolis, MN 55455}}
\centerline{\large \it Dept. Physics and Astronomy, Univ. New Mexico, Albuquerque, NM 87131}

\vspace{8pt}
\centerline{\large \bf Raymond Chan}
\centerline{\large \it Dept., Mathematics, Chinese Univ. of Hong Kong, Shatin, Hong Kong}

\vspace{8pt}
\centerline{\large \bf Mila Nikolova}
\centerline{\large \it ENS Cachan, Univ.  Paris-Saclay, 94235 Cachan Cedex, France }
\vspace{15pt}


\section*{ABSTRACT} 
\vspace{9pt}
{An optical imager that exploits off-center image rotation to encode both the lateral and depth coordinates of point sources in a single snapshot can perform 3D localization and tracking of space debris. When actively illuminated, unresolved space debris, which can be regarded as a swarm of point sources, can scatter a fraction of laser irradiance back into the imaging sensor. Determining the source locations and fluxes is a large-scale sparse 3D inverse problem, for which we have developed efficient and effective algorithms based on sparse recovery using non-convex optimization. Numerical simulations illustrate the efficiency and stability of the algorithms.}



\section{INTRODUCTION}
\label{sec:intro}  

We consider 3D localization and tracking of space debris at optical wavelengths by using a space-based telescope, which is an important and challenging task in space surveillance. Since the optical wavelength is much shorter than the radio wavelength, optical detection and localization is expected to attain far greater precision than the more commonly employed radar systems. However, the shorter field depth of optical imaging systems may limit their performance to a shorter range of distances. An integrated system consisting of a radar system for performing radio detection, localization, and ranging of space debris at larger distances, which cues in an optical system when debris reach shorter distances, may ultimately provide optimal performance for detecting and tracking debris at distances ranging from tens of kilometers down to hundreds of meters.

A stand-alone optical system based on the use of a light-sheet illumination and scattering concept \cite{englert2014optical}
for spotting debris within meters of a spacecraft has been proposed. 
A second system can localize all three coordinates of an unresolved, scattering debris \cite{hampf2015optical,wagner2016detection} by
utilizing either parallex between two observatories or a pulsed laser ranging system or a hybrid system. For parallex, two observatories receive debris scattered optical signals simultaneously. For the pulsed laser, the ranging system is coupled to a single imaging observatory. The hybrid system utilizes both approaches in which the laser pulse transmitted from one of the two observatories is received at time-gated single-photon detectors with good parallax information at both the observatories. 
 However, to the best of our knowledge there is no other proposal for a full 3D debris localization and tracking optical or optical-radar system working in the range of tens to hundreds of meters. 
 Prasad \cite{Internal_report2016prasad} has proposed 
the use of an optical imager that exploits off-center image rotation to encode in a single image snapshot both the range $z$ and transverse ($x,y$) coordinates of a swarm of unresolved sources such as 
small, sub-centimeter class space debris, which when actively illuminated can scatter a fraction of laser irradiance back into the imaging sensor.

Image data taken with a specially designed point spread function (PSF) that encodes, via a simple rotation, changing source distance can be employed to acquire a three dimensional (3D) field of unresolved sources like space debris. 
By imposing spiral phase retardation with a phase winding number that changes in regular integer steps from one annular zone to the next of an aperture-based phase mask, one can create an image of a point source that has an approximate rotational shape invariance {with changing source distance}, provided the zone radii have a square root dependence on their indices. Specifically, when the distance of the source from the aperture of such an imaging system changes, the off-center, shape-preserving PSF merely rotates by an amount roughly proportional to the source misfocus from the plane of best focus. 
The following general model based on the rotating PSF image describes the spatial distribution of image brightness for $M$ point sources {describes the observed 2D image}: 
\begin{equation}
	G (x,y) = \mathcal{N}\left(\sum_{i=1}^M \mathcal{H}_{{z_i}}(x-x_i,y-y_i)f_i + b\right), 
	\label{equ:forward_model}
\end{equation}
where  $\mathcal{N}$ is the noise operator and $b$ is the uniform background value. Here  
$\mathcal{H}_i(x-x_i,y-y_i)$ is the rotating PSF for the $i$-th point source of flux $f_i$ and  3D position coordinates $(x_i, y_i, z_i)$ with the depth information $z_i$ encoded in  $\mathcal{H}_i$, and $(x,y)$ is  the position in the image plane.

A simple approach to effect such PSF rotation, which was originally proposed by Prasad \cite{prasad2013rotating} in 2013, utilizes an annular phase mask design of spiral phases with winding numbers that are regularly spaced from one annular zone to the next. Such a mask can be easily mounted on a telescope. When actively illuminated by a laser, unresolved space debris, which can be regarded as a swarm of point sources, can scatter a fraction of the laser irradiance back into the imaging sensor. The technique is well suited to optically localize small, sub-centimeter class space debris, which we may call microdebris, at distances of hundreds of meters.

Following the Fourier optics model, the  incoherent PSF for a clear aperture containing a phase mask with optical phase retardation, $\psi(\mathbf{s})$, is given by 
$${\mathcal{H}_{z}}(\mathbf{s} ) = 
{1\over \pi}\left|\int P(\mathbf{u} )\mathrm{exp} \left[ \iota( 2\pi \mathbf{u}\cdot\mathbf{s} +  \zeta u^2 - \psi(\mathbf{u}))  \right] d \mathbf{u} \right|^2,
$$  
where 
{$\zeta = \frac{\pi  (l_0-z) R^2}{\lambda l_0 z}$  is defocus parameter.  {Here $\iota = \sqrt{-1} $ and $\l_0$, and $P(\mathbf{s} )$ denote the distance between the lens and the best focus point and the indicator function for the pupil of radius $R$, respectively, while} $\mathbf{s}$ {with polar coordinates $(s, \phi_{\mathbf{s}}) $} is a scaled version of the image-plane position vector, $\mathbf{r}$, namely $\mathbf{s} =  \frac{\mathbf{r}}{\lambda z_I/R} $.  
Here $\mathbf{r}$ is measured from the center of the geometric {(Gaussian)} image point located at $\mathbf{r}_I$. The pupil-plane position vector $\boldsymbol{\rho}$ is normalized by the pupil radius, $\mathbf{u} = \frac{\boldsymbol{\rho}}{R}$. 
For the single-lobe rotating PSF, $\psi(\mathbf{u}) $ is chosen to be the spiral phase
 profile defined as $$\psi(\mathbf{u}) = l\phi_{\mathbf{u}}, \ \ \text{for } \sqrt{\frac{l-1}{L}}\leq u\leq \sqrt{\frac{l}{L}}, \ l = 1,\cdot\cdot\cdot, L, $$  in which $L$ is the number of concentric annular zones in the phase mask. We evaluate (\ref{equ:A}) by using the fast Fourier transform.

We discuss here the problem of 3D localization of closely spaced point sources from simulated noisy image data obtained by using such a rotating-PSF imager. The localization problem is discretized on a cubical lattice where the coordinates and values of its nonzero entries represent the 3D locations and fluxes of the sources, respectively. Finding the locations and fluxes of a few point sources on a large lattice is evidently a large-scale sparse 3D inverse problem. For the Gaussian and Poisson statistical noise models, we describe the results of simulation using novel non-convex sparse optimization algorithms to extract both the 3D location coordinates and fluxes of individual debris particles from noisy rotating-PSF imagery. For the Gaussian noise case, which describes conventional CCD sensors operating at low per-pixel photon fluxes and large read-out noise, a continuous exact $\ell_0$ (CEL0) penalty term \cite{CEL02015soubies} added to a least-squares data fitting term constitutes an $\ell_0$-sparsity non-convex optimization protocol with promising results. For the Poisson noise case, which characterizes an EMCCD sensor operated in the photon-counting (PC) regime, we show that an iteratively reweighted $\ell_1$ (IRL1) algorithm based on the sum of a Kullback-Leibler $I$-divergence data fitting term and a novel non-convex penalty term \cite{mila2010nonconvex} performs well \cite{kl_nc}. Image data of the type we discuss here could be acquired by a combined active illumination - imaging system that can be mounted on a space asset in order to optically monitor its debris neighborhood. Further work involving snapshot multi-spectral imaging for material characterization and higher 3D resolution and localization of space microdebris via a sequence of snapshots is underway.

The rest of the paper is organized as follows. In Section \ref{sec:non-convex}, we propose non-convex optimization methods to solve the point source localization problem for both Gaussian and Poisson noise models. In Section \ref{sec:alg}, our non-convex optimization algorithms are developed. A new iterative scheme for estimating the flux values for the Poisson noise case is also proposed in this section. Numerical experiments, including comparisons with other optimization methods, are discussed in Section \ref{sec:numerical}.  Some concluding remarks are made in Section \ref{sec:Conclusions}.


\section{NON-CONVEX OPTIMIZATION MODELS FOR GAUSSIAN AND POISSON NOISE }\label{sec:non-convex}

Here, we  build  forward models for the problem based in part  on the approach developed in  \cite{Rice2016generalized}. 
In order to estimate the 3D locations of the point sources, we assume their distribution is approximated by a discrete lattice  $\mathcal{X}\in \mathds{R}^{m \times n \times d}$. 
The indices of the nonzero entries of $\mathcal{X}$ are the 3-dimensional locations of the point sources and the values at these entries   correspond to the fluxes, {\it i.e.}, the energy emitted by the illuminated point source. The 2D observed image $G \in \mathds{R}^{m \times n}$ can be represented as 
\begin{equation*}
	G = \mathcal{N}\left(\mathcal{T}(\mathcal{A} \ast \mathcal{X} )+b 1  \right),
\end{equation*}
where
 $b$ is background signal, $1$ is a matrix of $1$s of size the same as the size of $G$ and 
$\mathcal{N}$ is the noise operator. Here   $\mathcal{A} \ast \mathcal{X}$ is the convolution of $\mathcal{X}$ with the 3D PSF $\mathcal{A}$. This 3D PSF ${\mathcal{A}}$ is a cube which is constructed by a sequence of images with respect to different depths of the points. Each slice is the image corresponding to a point source at the origin in the $(x,y)$ plane and at depth $z$.  The dictionary $\mathcal{A}$ {is constructed} by sampling depths at regular intervals in the range, $\zeta_i\in [-\pi L, \ \pi L]$, over which the PSF performs one complete rotation about the geometric image center before it begins to break apart. The $i$-th slice of dictionary is {$\mathcal{H}_{z_i}$ with certain depth $z_i$}.  Here $\mathcal{T}$ is an operator for extracting the last slice of the cube $\mathcal{A} \ast \mathcal{X}$ since the observed information is a snapshot, and $\mathcal{N}$ is the  noise operator.

 In order to recover $\mathcal{X}$,  we need to solve a large-scale sparse 3D inverse problem given as follows: 

\begin{equation}\label{model}  
        \min_\mathcal{X} \mathcal{D} (\mathcal{T}(\mathcal{A}\ast \mathcal{X})+b, G) + \mathcal{R}(\mathcal{X}),
\end{equation}
where $\mathcal{R}(\mathcal{X})$ is {a regularization, or penalty,} term to approximate the $\ell_0$ pseudo-norm which {\it gives the number of nonzero entries} in $\mathcal{X}$.  Here $\mathcal{D}$ is a certain data-fitting term based on the noise model. 

In the following sections, we consider the Gaussian and Poisson noise cases.
For notation purposes we define $\ell_2 - \ell_k$  to denote the inverse problem (\ref{model}), where $\mathcal{D} =\ell_2$ denotes the least squares fitting term and $ \mathcal{R} = \ell_k $ denotes the regularization term, with $k = 0$ or $1$. We extend this notation to define $\mathcal{R} = $ CEL0  and  $\mathcal{R} = $ NC to denote specific non-convex regularization terms.


\subsection{$\ell_2$-CEL0 (Gaussian noise case)}\label{sec:cel0}
When conventional CCD sensors operate at low per-pixel photon fluxes with large read-out noise, the noise $\mathcal{N}$ can be described as Gaussian noise. The noise is data-independent, which leads to the use of  least squares for the data-fitting term, {\it i.e.,} $$\mathcal{D} (\mathcal{T}(\mathcal{A}\ast \mathcal{X})+b, G):= \frac{1}{2} \left\|\mathcal{T}(\mathcal{A}\ast \mathcal{X})+b-G\right\|_F^2, $$ 
where  $\|X\|_F$ is the Frobenius norm of $X$, which is equal to the $\ell_2$ norm of the vectorized $X$. 
For the regularization term, we choose the continuous exact $\ell_0$ (CEL0) penalty, as described in  \cite{CEL02015soubies}.  It is a non-convex term approaching the $\ell_0$ 
 norm for linear least squares data fitting problems. $\mathcal{R}(\mathcal{X})$ is constructed as 
 {
 $$\mathcal{R}(\mathcal{X}):=\Phi_{\mathrm{CEL0}}(\mathcal{X})=\sum_{u,v,w = 1}^{m,n,d}\phi(\|\mathcal{T}(\mathcal{A} \ast \delta_{uvw} )\|, \mu,; \mathcal{X}_{uvw}), $$
  where $\phi(a, \mu,; u) = \mu  - \frac{a^2}{2}\left(|u|-\frac{\sqrt{2\mu}}{a}\right)^2 \mathds{1}_{\left\{|u| \leq \frac{\sqrt{2 \mu  }}{a} \right\}},  \quad 
\mathds{1}_{\{ u \in E\}} := \begin{cases}
	1 & \text{if} \ u\in E;\\
	0 & \text{others}.
\end{cases}   $ (see \Cref{fig:curve_regul}(a)
and $\delta_{uvw}$ is a 3D tensor whose only nonzero entry is at $(u, v, w)$ with value 1. Here $\mu $ is the regularization parameter and $u \in  1, \cdots, m, v \in  1,2, \cdots, n, w \in  1,\cdots, d. $ 
\begin{figure}[htbp]
\centering
\subfloat[$\phi(a,\mu;u)$]{\includegraphics[width=0.39\textwidth, height=0.22\textheight]{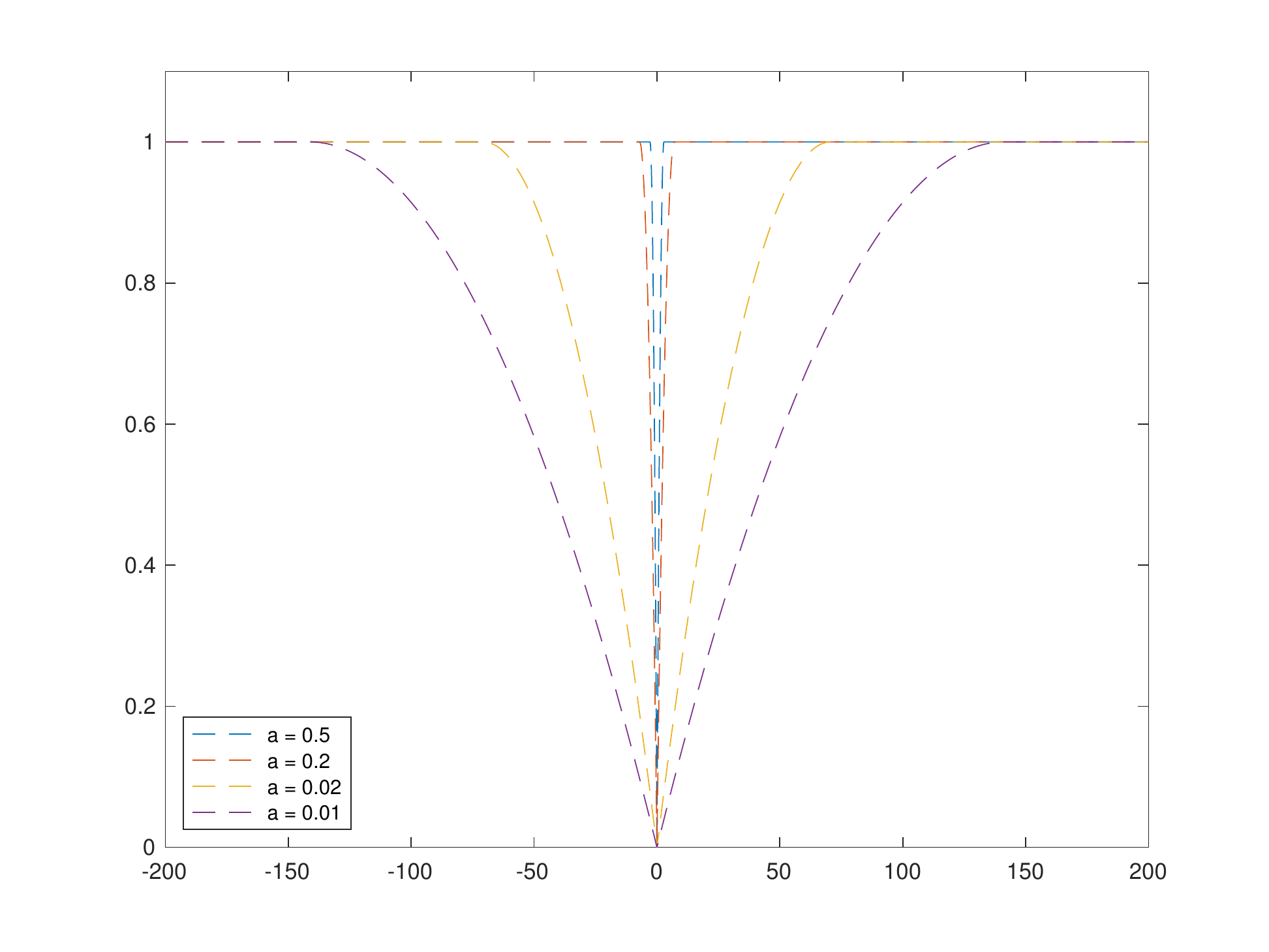}}
\subfloat[$\theta(a;u)$]{\includegraphics[width=0.39\textwidth, height=0.22\textheight]{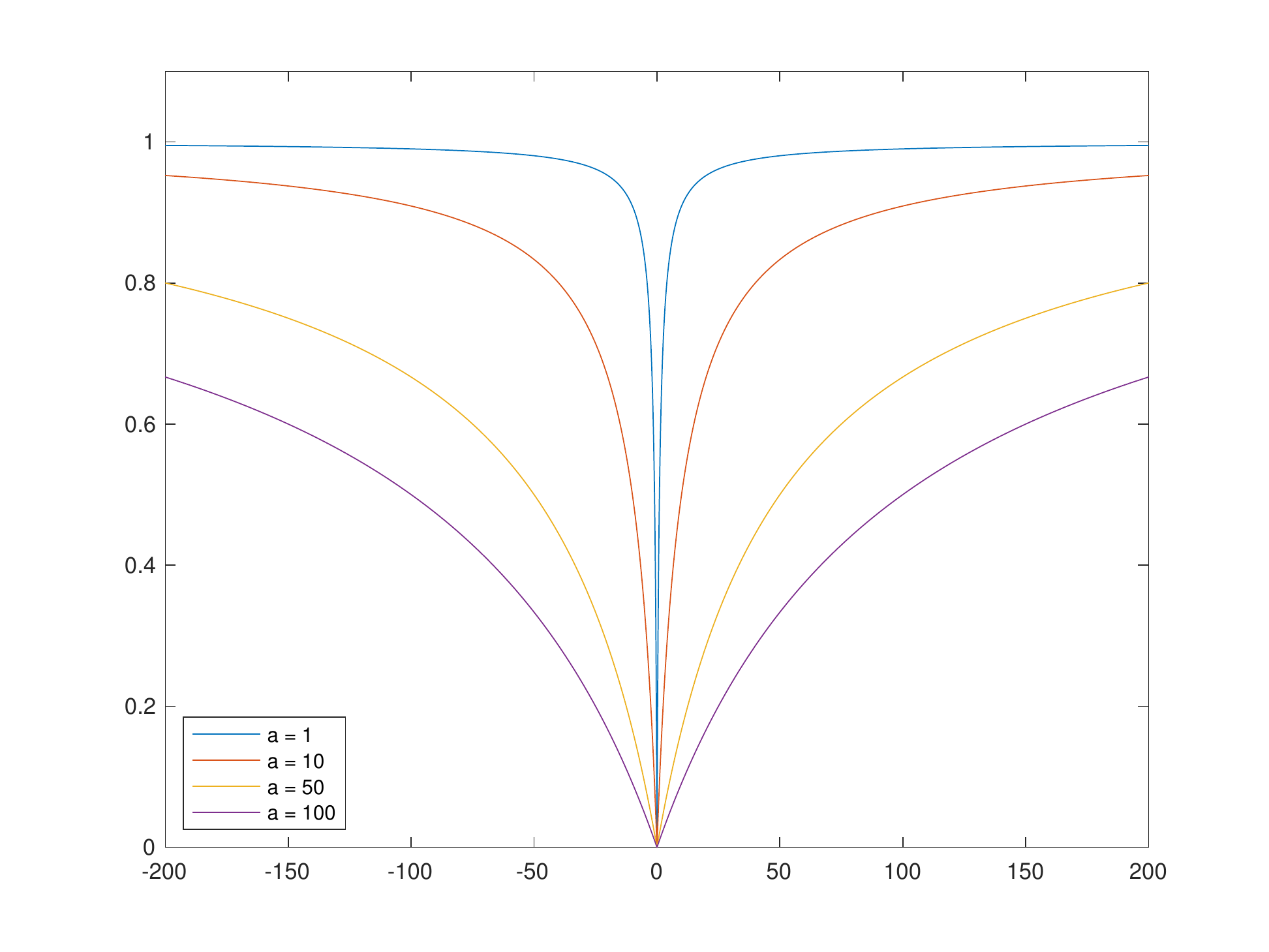}}
\caption{Non-convex regularization approaching to $\ell_0$ pseudo-norm with different value of $a$. Here the left subfigure is the function for $\ell_2$-CEL0 in  with $\mu=1$, the right subfigure is the function for KL-NC in Poisson noise case.  } 
\label{fig:curve_regul}
\end{figure}

The minimization problem amounts to 
	 \begin{equation}
	 	\min\limits_{\mathcal{X}\geq 0 }\left\{  \frac{1}{2} \left\|\mathcal{T}(\mathcal{A}\ast \mathcal{X})+b-G\right\|_F^2 +\sum_{u,v,w = 1}^{m,n,d}\phi(\|\mathcal{T}(\mathcal{A} \ast \delta_{uvw} )\|, \mu,; \mathcal{X}_{uvw}) \right\} . \label{equ:min_fun_gaussian}
	 \end{equation}
	 }To emphasize that our non-convex optimization model is based on the use of a least squares data fitting ($\ell_2$) term and the CEL0 regularization term, we designate our optimization model \eqref{equ:min_fun_gaussian} as $\ell_2$-CEL0.

We note that $\ell_2$-CEL0 has many good properties and it does not need any strict requirements on {the least squares data-fitting term}.
{The global minimizers of the $\ell_0$ penalty model with a least squares data-fitting term ($\ell_2$-$\ell_0$)  are contained in the set of global minimizers of $\ell_2$-CEL0 \eqref{equ:min_fun_gaussian}. A minimizer of \eqref{equ:min_fun_gaussian} can be transformed into a minimizer of  $\ell_2$-$\ell_0$. Moreover, some local minimizers of $\ell_2$-$\ell_0$ are not critical points of $\ell_2$-CEL0, which means $\ell_2$-CEL0 can avoid some local minimizers of $\ell_2$-$\ell_0$ .
} 



\subsection{KL-NC (Poisson noise case)}\label{klnc}
 Here we consider the Poisson noise case for which the data fitting term is the $I$-divergence. This term is is also known as Kullback-Leibler (KL) divergence \cite{poisson_formula}, and can be expressed as follows for our case: 
	   \begin{equation*}
	   		\mathcal{D} (\mathcal{T}(\mathcal{A}\ast \mathcal{X})+b, g) := D_{KL}(\mathcal{T}(\mathcal{A}\ast \mathcal{X})+b, G),
	   \end{equation*}
	   where $D_{KL}(z,g) = \langle g, \ln \frac{g}{z} \rangle + \langle 1, z-g \rangle. $
For the regularization term, {the good properties of the CEL0 penalty term fail as the deta-fitting term is no longer least squares, }therefore, we consider a new non-convex 
function 
(see \cite{mila2008nonconvex,mila2010nonconvex,mila2013nonconvex,mila2015nonconvex}), using specifically 
\begin{equation*}
	\mathcal{R}(\mathcal{X}):= \mu \sum_{i,j,k = 1}^{m,n,d} \theta(a;\mathcal{X}_{ijk}) = \mu \sum_{i,j,k = 1}^{m,n,d} \frac{|\mathcal{X}_{ijk}|}{a+|\mathcal{X}_{ijk}|},
\end{equation*}
where $a$ is fixed and 
determines the degree of non-convexity. (see \Cref{fig:curve_regul}(b))
Thus, the minimization problem amounts to 
	 \begin{equation}
	 	\min\limits_{\mathcal{X}\geq 0 }\left\{  \left\langle 1,  \mathcal{T}(\mathcal{A} \ast \mathcal{X})- G \ln(\mathcal{T}(\mathcal{A} \ast \mathcal{X})+b \,1) \right\rangle + \mu\sum_{i,j,k = 1}^{m,n,d} \frac{|\mathcal{X}_{ijk}|}{a+|\mathcal{X}_{ijk}|}\right\}. \label{equ:min_fun_poisson}
	 \end{equation}
Here, $\theta(a;t) = \lim\limits_{\epsilon \to 1} \theta_\epsilon(a;t), $ where $\theta_\epsilon(t) = \frac{|t|}{a+ \epsilon |t|}. $ Since $\theta_\epsilon(t)$ represents the $\ell_1$ norm when $\epsilon = 0$, we observe that the process of increasing non-convexity as $\epsilon$ increases from 0 to 1. 

{\bf Remark:}	To emphasize that our non-convex optimization model is based on the use of a KL data fitting (KL) term and a 
non-convex (NC) regularization term, we designate our  optimization model \eqref{equ:min_fun_poisson} as KL-NC.

\section{ DEVELOPMENT OF OUR ALGORITHMS  }\label{sec:alg}

 Note that our optimization models for both the Gaussian and Poission noise cases are  non-convex, due to the regularization terms. 
We first consider an iterative reweighted $\ell_1$ algorithm (IRL1) \cite{IRL1_2015} to solve the optimization problems. This is a majorization-minimization method which solves a series of convex optimization problem with a weighted-$\ell_1$ regularization term. 
 It considers the problem (see Algorithm 3, in \cite{IRL1_2015})
\begin{equation*}
	\min_{x\in X} F(x):=F_1(x)+ F_2(G(x)),
\end{equation*}
 where $X$ is the constraint set. $F$ is a lower semicontinuous (lsc) function, extended, real-valued, proper, while $F_1$ is proper, lower-semicontinous, and convex and $F_2$ is coordinatewise nondecreasing, {\it i.e.} $F_2(x)\leq F_2(x+t e_i)$ with $x, x+t e_i\in G(X)$ and $t >0,$ where  $e_i$ is the $i$-th canonical basis unit vector. The function $F_2$ is concave on $G(X)$.   The IRL1 iterative scheme \cite[Algorithm 3]{IRL1_2015} is 
 \begin{equation*}
 	\begin{cases}
 		w^l = \partial F_2(y), \ y = G(x^l), \\
 		x^{l+1} = \argmin\limits_{x\in X} \left\{F_1(x) + \langle w^l, G(x)\rangle \right\},
 	\end{cases}
 \end{equation*}
where $\partial$ stands for subdifferential.

For the Gaussian noise case \eqref{equ:min_fun_gaussian}, we choose
\begin{equation*}
	\begin{split}
		F_1(\mathcal{X}) = & \frac{1}{2} \left\|\mathcal{T}(\mathcal{A} \ast \mathcal{X} )+b 1 -G\right\|_F^2;   \\
		F_2(\mathcal{X}) = & \  \mu  - \frac{\|a_i\|^2}{2}\left(\mathcal{X}_{ijk}-\frac{\sqrt{2\mu}}{\|a_i\|}\right)^2 \mathds{1}_{\left\{\mathcal{X}_{ijk} \leq \frac{\sqrt{2 \mu  }}{\|a_i\|} \right\}}; \\
		G(\mathcal{X}) = & \  |\mathcal{X}|; \\
		X = & \ \{\mathcal{X} \ | \ \mathcal{X}_{ijk} \geq 0 \text{ for all } i,j,k  \}. 
	\end{split}
\end{equation*}

{\bf Remark: }The minimization problem in each iteration of IRL1 is a weighted  $\ell_1$ model with nonnegative constraints. In  \cite{Rice2016generalized}, the $\ell_1$ model without nonnegative constraints is solved by the alternating direction method of multipliers (ADMM).

For the Poisson noise case \eqref{equ:min_fun_poisson}, we can choose the same $G(\mathcal{X})$ and $X$ as Gaussian noise case, but $F_1$ and $F_2$ are as follows:
\begin{equation*}
	\begin{split}
		F_1(\mathcal{X}) = & \ \langle 1, \  \mathcal{T}(\mathcal{A} \ast \mathcal{X})- G \log(\mathcal{T}(\mathcal{A} \ast \mathcal{X})+b \,1) \rangle;  \\
		F_2(\mathcal{X}) = & \  \mu \sum_{i,j,k = 1}^{m,n,d} \frac{\mathcal{X}_{ijk}}{a+\mathcal{X}_{ijk}}. \\
	\end{split}
\end{equation*}
Therefore, we compute the partial derivative of $w^l$ and get  $w^l_{ijk} =\frac{a \mu } { \left(a + \hat{\mathcal{X}}_{ijk}^l  \right)^2 }, \quad \forall i,j,k$. Here $w^l\neq0$ is finite, since $a, \mu \neq 0$, and all $\mathcal{X}_{ijk}\geq 0$ owning to the constraint $X$.   
According to \cite{mila2008nonconvex,poisson_marcia}, these terms satisfy the requirements of the algorithm.

For real data, the point sources may be not on a grid, which means the discrete model may not be accurate. In order to avoid missing point sources, the regularization parameter $\mu$ is kept small, which can potentially lead to over-fitting. 
Our optimization solution generally contains tightly clustered point sources, so we need to regard any such cluster of point sources as a single point source.
The same phenomenon has been observed in \cite{Rice2016generalized,FALCON2014,clustered2012fasterstorm}. 
We apply a post-processing approach following
\cite{clustered2012fasterstorm}.  The method is based on the well-defined tolerance distance for recognizing clustered neighbors. 
We compute the centroid of each cluster, which we regard as a single point source. 

\subsection{Flux estimation}\label{sec:flux}
In the Poisson noise case, 
our numerical results show that the flux values are generally underestimated. In \cite{Rice2016generalized,FALCON2014},  least squares fitting is used for improving the resolution as well as updating the corresponding fluxes. However, our problem is not {a Gaussian-noise problem, and in fact an additive} Poisson noise as used in these paper. Our Poisson noise is data-dependent, which cannot {constrain, with least squares, the observed data to match the regenerated image $\mathcal{T}(\mathcal{A}\ast \mathcal{X^\ast})$, in which $\mathcal{X}^\ast$ is the scene estimate}.
Our aim is thus to estimate the source fluxes from the KL data-fitting term appropriate to the Poisson noise model
after the source 3D positions have already been accurately estimated. 

Let the PSF corresponding to the $i$-th source
be arranged as the column vector $\mathbf{h}_{i}$. The stacking of the $P$ column vectors in the same sequence as
the source labels for the $P$ sources then defines a system PSF matrix $H$, with $H= \left[\mathbf{h}_1, \mathbf{h}_2, \cdots, \mathbf{h}_P\right]\in \mathds{R}^{K\times P}, $ where $K$ is the total number of pixels in the vectorized data array, so $K=mn. $
The vectorized observed image is denoted by  $\mathbf{g} \in \mathds{R}^{ K \times 1}$.  The uniform background is denoted as  the vector $b\mathbf{1}$ with  $\mathbf{1}\in \mathds{R}^{K\times 1}$.
The flux vector is denoted as $\mathbf{f}\in \mathds{R}^{P\times 1}$.  Here the problem is overdetermined, {\it i.e.,}  the number of point sources $P$ is much smaller than the number of available data $K = mn$.   Therefore we need to do some refinement of the estimates by minimizing directly data fitting term. 
 Since the negative log-likelihood function for the Poisson model, up to certain data dependent terms, 
is simply the KL divergence function,
\begin{equation*} 
	D_{KL}(H \mathbf{f} +b \mathbf{1}, \mathbf{g}) = \left\langle \mathbf{1}, H \mathbf{f}{+b \mathbf{1}} - \mathbf{g} \log(H \mathbf{f} + b \mathbf{1})  \right\rangle, 
\end{equation*}
its minimization with respect to  the flux vector $\mathbf{f}$, performed
by setting the gradient of $D_{KL}$  with respect  to $\mathbf{f}$ (see \cite{poisson_marcia}) zero. This yields the nonlinear relation
\begin{equation}
\label{e2}
\begin{split}
\nabla D_{KL}(H \mathbf{f} +b \mathbf{1}, \mathbf{g}) = & H^T \mathbf{1} - \sum\limits_{i = 1}^K \frac{\mathbf{g}_i}{\mathbf{e}_i^T (H \mathbf{f} + b \mathbf{1})} H^T \mathbf{e}_i \\
= & \sum\limits_{i = 1}^K\frac{\mathbf{e}_i^T\left( H \mathbf{f} + b \mathbf{1} -\mathbf{g}\right)}{\mathbf{e}_i^T (H \mathbf{f} + b \mathbf{1})} H^T \mathbf{e}_i = 0,
\end{split}
\end{equation} 
{where $\mathbf{e}_i$ is the $i$-th canonical basis unit vector.
Consider now an iterative solution of \eqref{e2}, {which may be expressed as 
the equality} 
%
\begin{equation}
	\label{equ:formula_f}
	\mathbf{f} = \mathbf{f}_G +  \mathcal{K}(\mathbf{f}), 
\end{equation}
where $\mathcal{K}(\mathbf{f}) = \sum\limits_{i = 1}^K\frac{\mathbf{e}_i^T\left( H \mathbf{f} + b \, 1-\mathbf{g} \right)\mathbf{e}_i^TH \mathbf{f}}{\mathbf{e}_i^T (H \mathbf{f} + b \mathbf{1})} H^+ \mathbf{e}_i $ and $H^{+} = (H^T H)^{-1}H^T$. Here $\mathbf{f}_G = H^{+}(\mathbf{g}-b \mathbf{1} )$ is the solution corresponding to the Gaussian noise model. 
This  suggests the following fixed point iterative scheme
\begin{equation}
	\mathbf{f}^{n+1} = \mathbf{f}_G + \mathcal{K}(\mathbf{f}^{n}),  \quad n = 1, 2, \cdots
	\label{iterative_scheme_flux}
\end{equation}
}
for estimating the flux.



\section{NUMERICAL RESULTS}\label{sec:numerical}

In this section, we apply our optimization  approaches to solving  simulated rotating PSF problems for point source localization and compare them to some other optimization methods. The codes of our algorithm and the others with which we compared our method were written in $\mathrm{MATLAB \ 9.0 \ (R2016a)}, $ and all the numerical experiments were conducted on a typical personal computer with a standard CPU (Intel i7-6700, 3.4GHz). 

The fidelity of localization is assessed in terms of the {\bf recall rate}, defined as the {\it ratio of the number of identified true positive point sources over the number of true positive point sources}, and the {\bf precision rate}, defined as the {\it ratio of the number of identified true positive point sources over the number of all point sources} obtained by the algorithm; see \cite{Book_SR_micro2017}. 

To distinguish  true positives from false positives for the estimated point sources, we need to determine the minimum total distance between thm and true point sources. Here all 2D simulated observed images are described by 96-by-96 matrices. We set the number of zones of the spiral phase mask responsible for the rotating PSF at $L=7$  and the aperture-plane side length as 4 which sets the pixel resolution in the 2D image (FFT) plane as 1/4 in units of $\lambda z_I/R$. 
The dictionary corresponding to our discretized 3D space contains 21 slices in the axial direction, with the corresponding values of the defocus parameter, $\zeta$, distributed uniformly over the range, $[-21, \  21]$.  
According to the Abbe-Rayleigh resolution criterion,  two point sources that are within $(1/2)\lambda z_I/R$ of each other and lying in the same transverse plane cannot be separated in the limit of low intensities.
In view of this criterion and our choice of the aperture-plane side length and if we assume conservatively that our algorithm does not yield any significant super-resolution, we must regard two point sources that  are within 2 image pixel units of each other as a single point source.
Analogously, two sources along the same line of sight ({\it i.e.,} with the same $x,y$ coordinates) that are axially separated from each other within a single unit of $\zeta$  must also be regarded as a single point source. 
  
As for real problems, our simulation does not assume that the point sources are on the grid points. Rather, a number of point sources are randomly generated in a 3D continuous image space with certain fluxes. We consider a variety of source densities, from 5 point sources to 40 point sources in the same size space.  For each density, we randomly generate 20 observed images and use them for training the parameters in our algorithm, and then test 50 simulated images with the well-selected parameters. The number of photons emitted by each point source follows a Poisson distribution with mean of 2000 photons. 


 {For adding Gaussian noise, we use} the MATLAB command 
  \begin{equation*}
  	\verb|G = I0  + b + sigma*randn(Np)|,
  \end{equation*}
where $\verb|b|$ is the uniform background noise which we set to a typical value 5. Here, $\verb|I0|$ is the 2D original image formed by  adding all the images of the point sources {without noise}, and  $\verb|Np| = 96$ is the size of the images. The noise level is denoted as $\verb|sigma|$. We choose  $\verb|sigma|$ to be   10\% of the highest pixel value in original image $\verb|I0|$. Here, \verb|randn| is the $\mathrm{MATLAB}$ command for  the Gaussian distribution with the mean as 0 and standard deviation as 1. 

For the Poisson noise case, we {\it apply Poisson noise not as additive noise as done in \cite{Rice2016generalized},  but rather as data-dependent Poisson noise} by using the MATLAB command 
  \begin{equation*}
  	\verb|G = poissrnd(I0+b)|,
  \end{equation*}
where  \verb|poissrnd| is the $\mathrm{MATLAB}$ command whose input is the mean of the Poisson distribution. 

\subsection{3D localizations for the Gaussian noise case}

In this subsection, we consider Gaussian noise and test our CEL0 based algorithm for several point-source densities. \Cref{fig:15-30_cel0} gives an example of 30 point sources. 


\begin{figure}[htbp]
\centering
\subfloat[Observed image]{\includegraphics[width=0.3\textwidth]{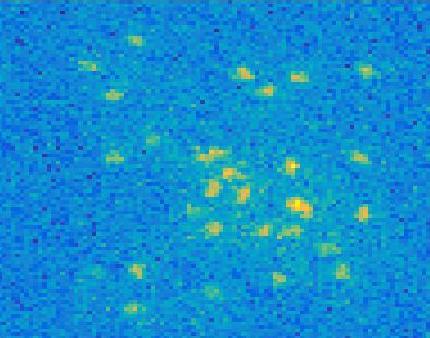}}
\hspace{0.01mm}
\subfloat[Estimated locations in 2D]{\includegraphics[width=0.3\textwidth]{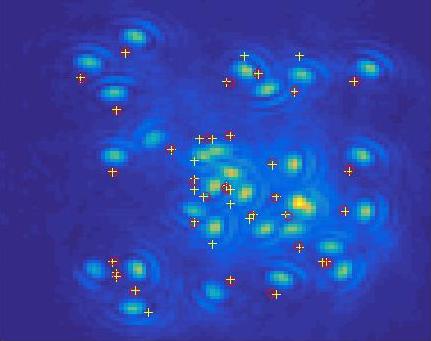}}
\hspace{0.01mm}
\subfloat[Estimated locations in 3D]{\includegraphics[width=0.3\textwidth]{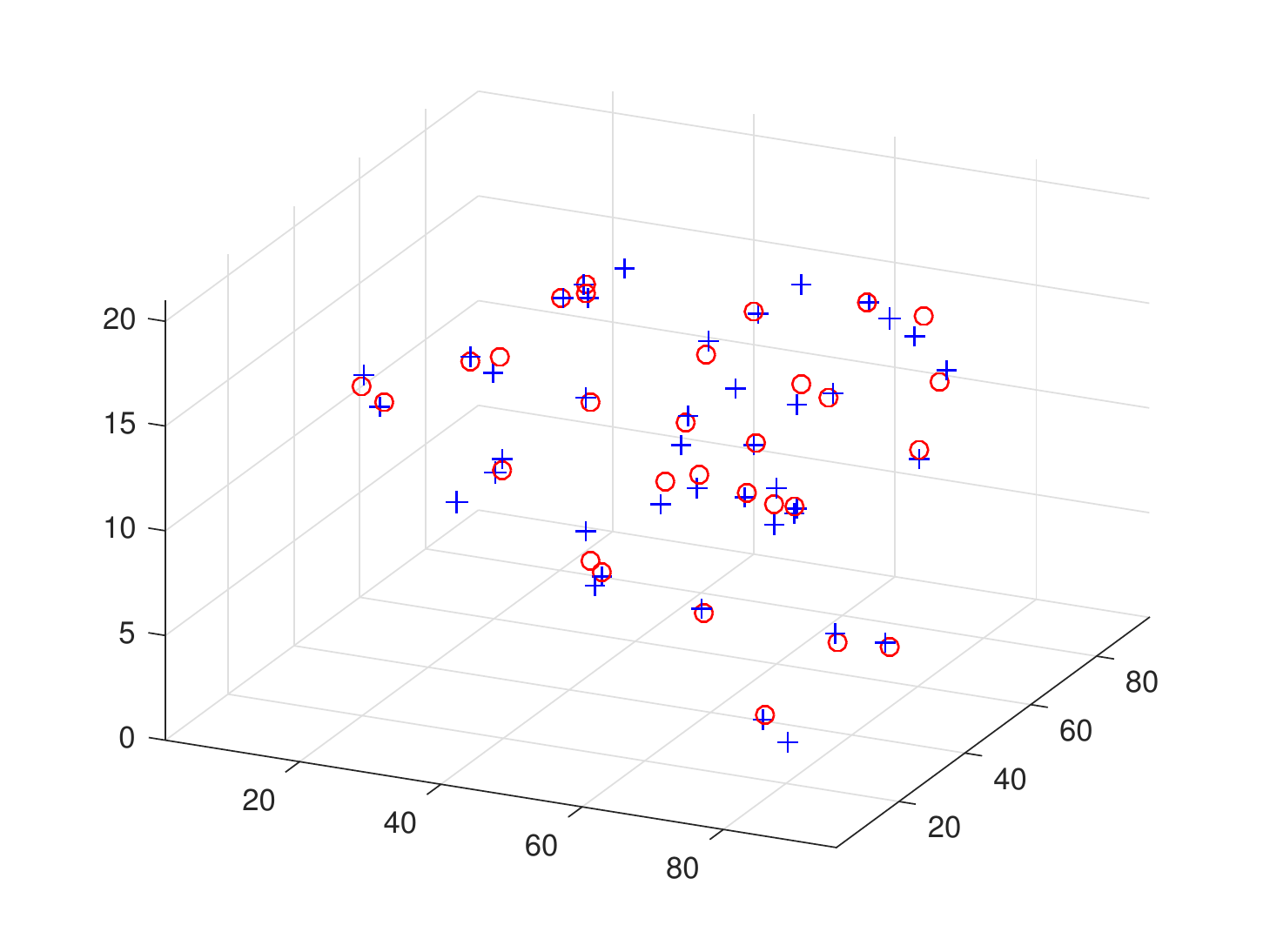}}
\caption{Gaussian noise case: Localizations for the 30 point sources case. ``$\circ$" denotes the location of the ground truth point source  and ``+"  the  location of the estimated point source. }\label{fig:15-30_cel0}
\end{figure}

  
In \Cref{fig:15-30_cel0},  many PSF  images are overlapping whose corresponding point sources are very close. Our algorithm estimates the clusters of these point sources but estimates more point sources than their ground-truth number. 
 

\begin{figure}[htbp]
\centering
\subfloat[$\ell_2$-$\ell_1$]{\includegraphics[width=0.39\textwidth, height=0.22\textheight]{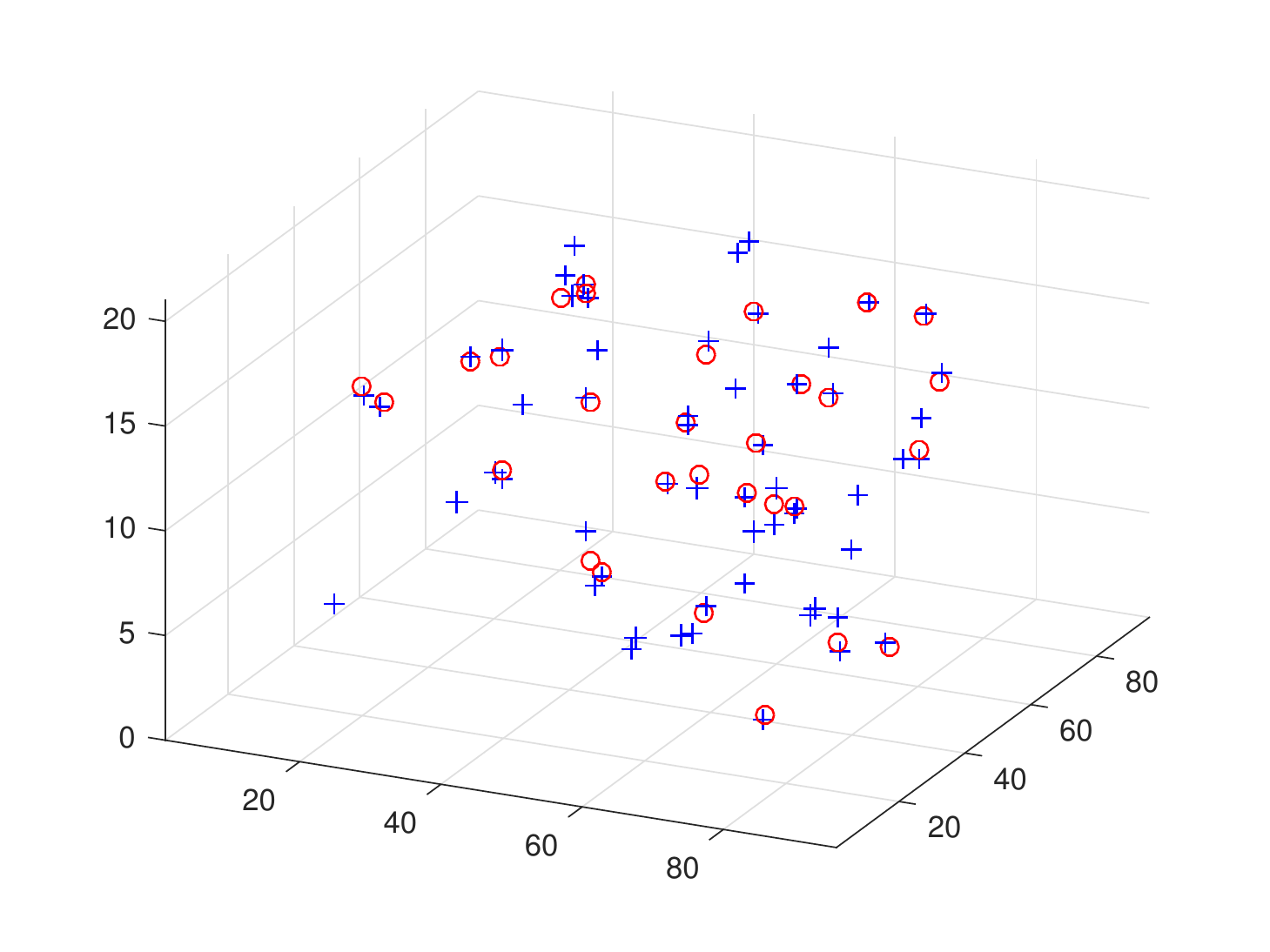}}
\subfloat[$\ell_2$-CEL0]{\includegraphics[width=0.39\textwidth, height=0.22\textheight]{images//30x_cel0.pdf}}
\caption{Gaussian noise case: Localizations from 2 algorithms (30 point sources). In (a), (b) and (c), ``$\circ$" denotes the location of the ground truth point source  and ``+"  the  location of the estimated point source.  } 
\label{fig:compare_distr_30 cel0}
\end{figure}

Next, we compare our algorithm with  $\ell_2$-$\ell_1$ (least squares fitting term with $\ell_1$ regularization model). In \Cref{fig:compare_distr_30 cel0},  we again consider the 30 point sources case.  We see that $\ell_2$-$\ell_1$ has more false positives than our algorithm although it detects all the ground truth point sources.  

For more comparison, we test 50 different random images and compute the average of recall and precision rate in each density case for both algorithms; see 
\Cref{Tab: cel0}.

\begin{table}[htbp]
\begin{center}
\caption{Gaussian noise case: Comparisons of  $\ell_2$-$\ell_1$ with our $\ell_2$-CEL0.  All the results are with post-processing.} 
\begin{tabular}{c|ccc|ccc}
\hline  & \multicolumn{3}{c}{$\ell_2$-$\ell_1$} & \multicolumn{3}{c}{$\ell_2$-CEL0}\\
\hline No.  Sources & Recall &  Prec. & Time & Recall & Prec. & Time  \\

\hline 5   & 95.60\% & 72.41\% & {\bf20.27} & {\bf98.00}\% & {\bf83.19}\% & 20.86\\
\hline 10  & 94.80\% & 64.04\% & {\bf19.99} & {\bf95.80}\% & {\bf79.72}\% & 20.93\\
\hline 15  & 90.80\%& 61.68\%  & {\bf20.09} &{\bf 93.20}\% &{\bf77.68}\% & 20.92\\
\hline 20  & 86.60\%& 57.72\% &  {\bf20.24} & {\bf 89.30}\% &{\bf72.12}\% & 20.25\\
\hline 30  & {\bf88.80}\% & 47.51\% & {\bf19.97} &87.20\% &{\bf58.77}\% & 21.12 \\
\hline 40  &  {\bf 81.50}\% & 42.03\% & {\bf19.95} & 77.40\% &{\bf52.87}\% & 21.09 \\
\hline
\end{tabular}\label{Tab: cel0}
\medskip
\end{center}
\end{table}

In \Cref{Tab: cel0}, our algorithm is better than $\ell_2$-$\ell_1$ almost for all cases especially in precision rate. For example, in 5 to 15 point sources case, the precision rate in our algorithm has over 10\% higher than the one in $\ell_2$-$\ell_1$.  In the high-density cases, like those with 30 and 40 sources, both methods have more than 5 false positives. We mitigate the latter by further post-processing based on machine learning technique, as in \cite{Rice2016generalized}. Here we must emphasize the advantage of our algorithm as providing a better initial guess than $\ell_2$-$\ell_1$ with similar cost time. {We set the maximum number of iterations for $\ell_2$-$\ell_1$ at 800, which guaranteed its convergence, and for CEL0 regularization, we set the maximum number of inner and outer iterations at 400 and 2, respectively.}

\subsection{3D localizations for the Poisson noise case}


\Cref{fig:15-30} shows another instance of 30 point sources, but for the case of Poisson noise and with many overlapping rotating PSF images. Such overlap in the presence of {data-dependent} Poisson noise makes the problem very difficult. The number and 3D locations of point sources is not easily obtained from observation. In this specific case, our algorithm still identifies all the true point sources correctly, but produces 9 false positives. From \Cref{fig:15-30}(b), we can see that these false positives come from the serious PSF overlapping.  

\begin{figure}[htbp]
\centering
\subfloat[Observed image]{\includegraphics[width=0.3\textwidth]{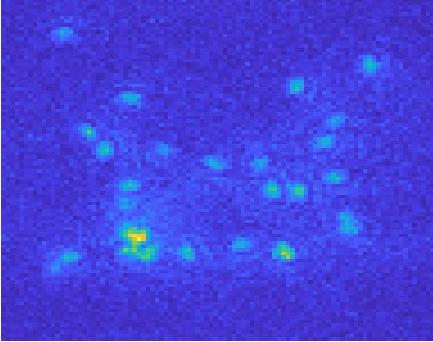}}
\hspace{0.01mm}
\subfloat[Estimated locations in 2D]{\includegraphics[width=0.3\textwidth]{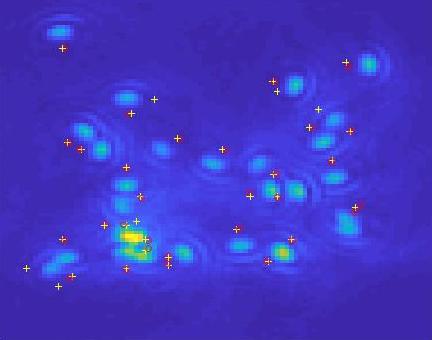}}
\hspace{0.01mm}
\subfloat[Estimated locations in 3D]{\includegraphics[width=0.3\textwidth]{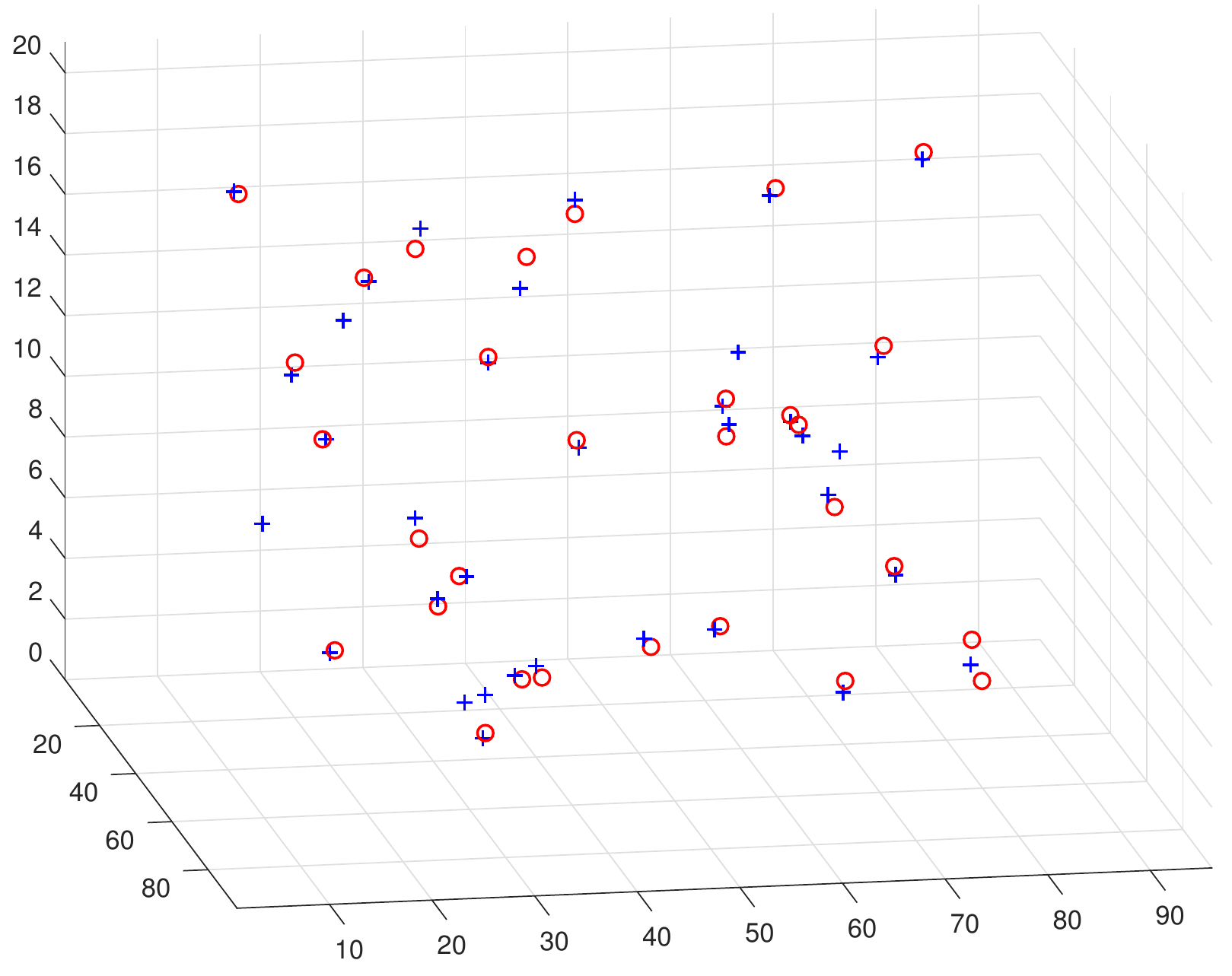}}
\caption{Poisson noise case: Localizations for the 30 point sources case. ``$\circ$" is the location of the ground truth point source  and ``+" is the  location of the estimated point source.  }\label{fig:15-30}
\end{figure}


Next, we compare our model with three other optimization models: KL-$\ell_1$ (KL data fitting with $\ell_1$ regularization); 
 $\ell_2$-$\ell_1$ (least squares fitting term with $\ell_1$ regularization) and $\ell_2$-NC (least squares fitting term with non-convex regularization model). For all these comparisons, we do the same post-processing and estimation of flux values after solving the corresponding optimization problem.

 Both the initial guesses of $\mathcal{X}$ and $\mathcal{U}_0$ are set as 0 for all these methods. In order to do the comparison, we plot the localizations for the four optimization models as well as the ground truth in the same space; see \Cref{fig:compare_distr_30} which correspond to the case of 30 point sources. From  \Cref{fig:compare_distr_30}, we see the overfitting of the $\ell_1$ regularization models (KL-$\ell_1$ and $\ell_2$-$\ell_1$). Before post-processing, the localizations of these two algorithms spread out the PSFs a lot and have many false positives. After post-processing, both algorithms are improved, especially KL-$\ell_1$. However, in comparison to the non-convex regularization (KL-NC and $\ell_1$-NC), they still have many more false positives. Among the four algorithms, our approach (KL-NC) performs the best in terms of the recall and precision rates.
  

\begin{figure}[htbp]
\centering
\subfloat[Without post-processing ($\ell_1$ regularization)]{\includegraphics[width=0.42\textwidth, height=0.25\textheight]{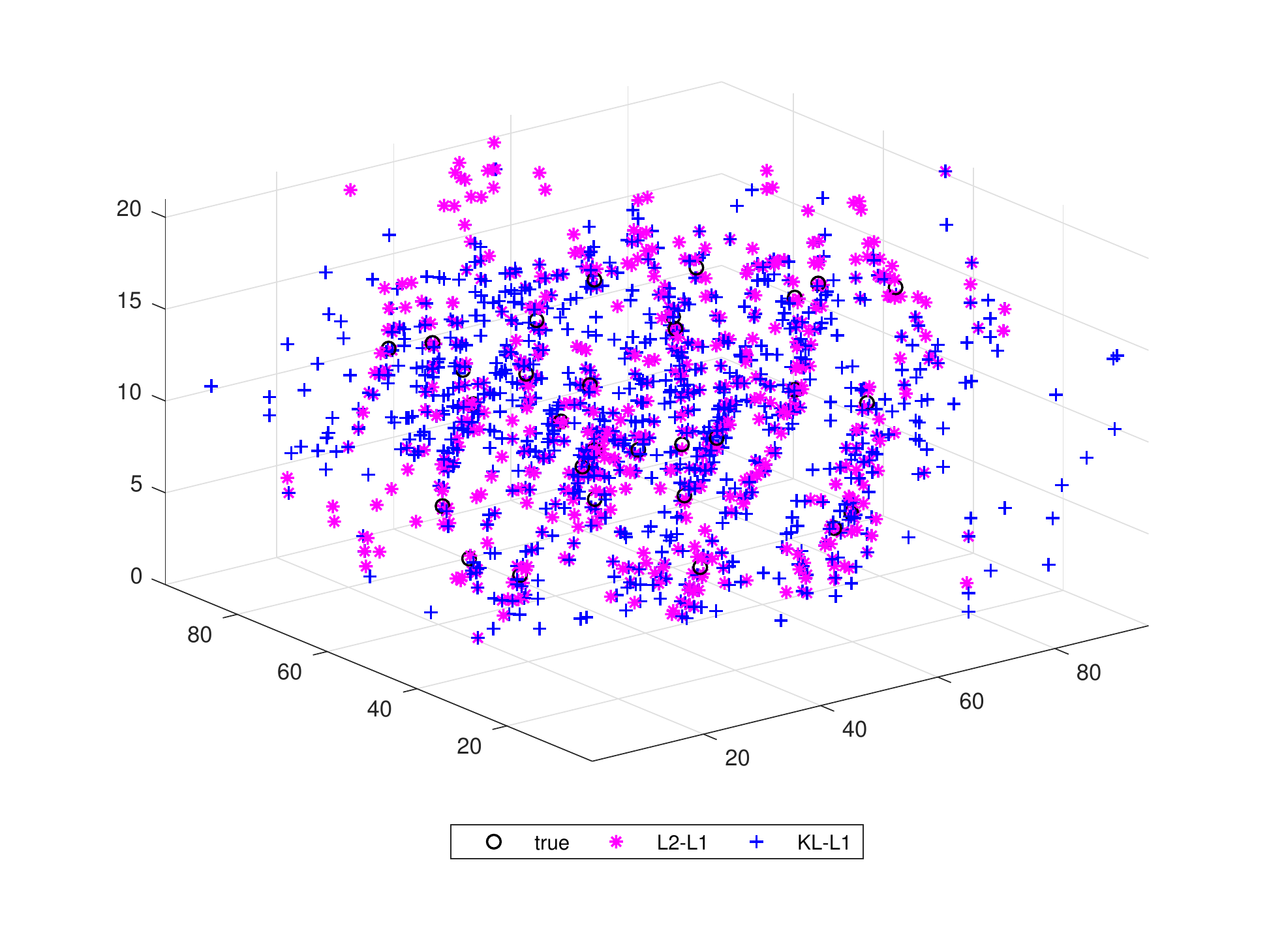}}
\subfloat[Without post-processing (non-convex regularization)]{\includegraphics[width=0.42\textwidth, height=0.25\textheight]{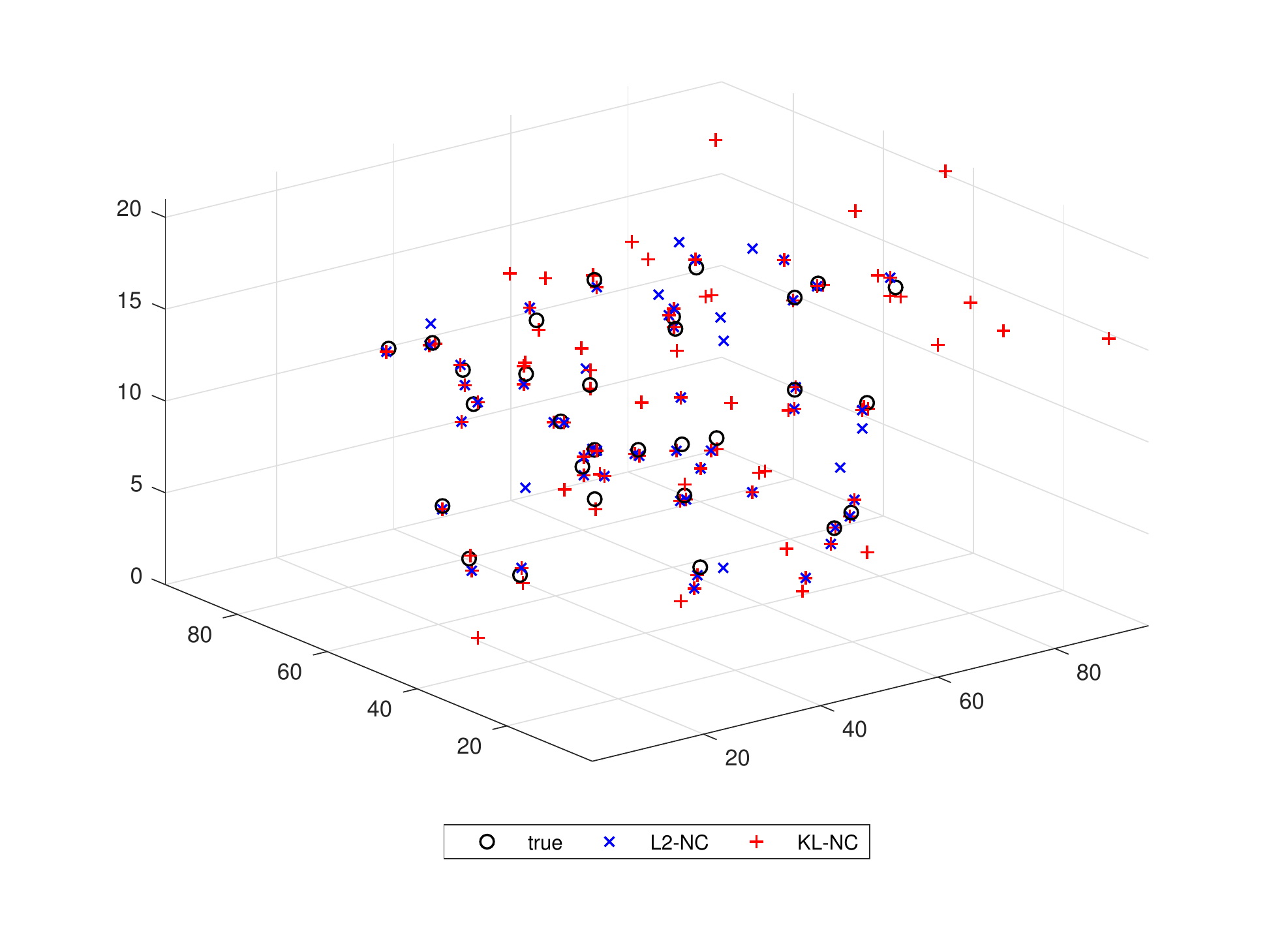}}\\
\subfloat[With post-processing ($\ell_1$ regularization)]{\includegraphics[width=0.42\textwidth, height=0.25\textheight]{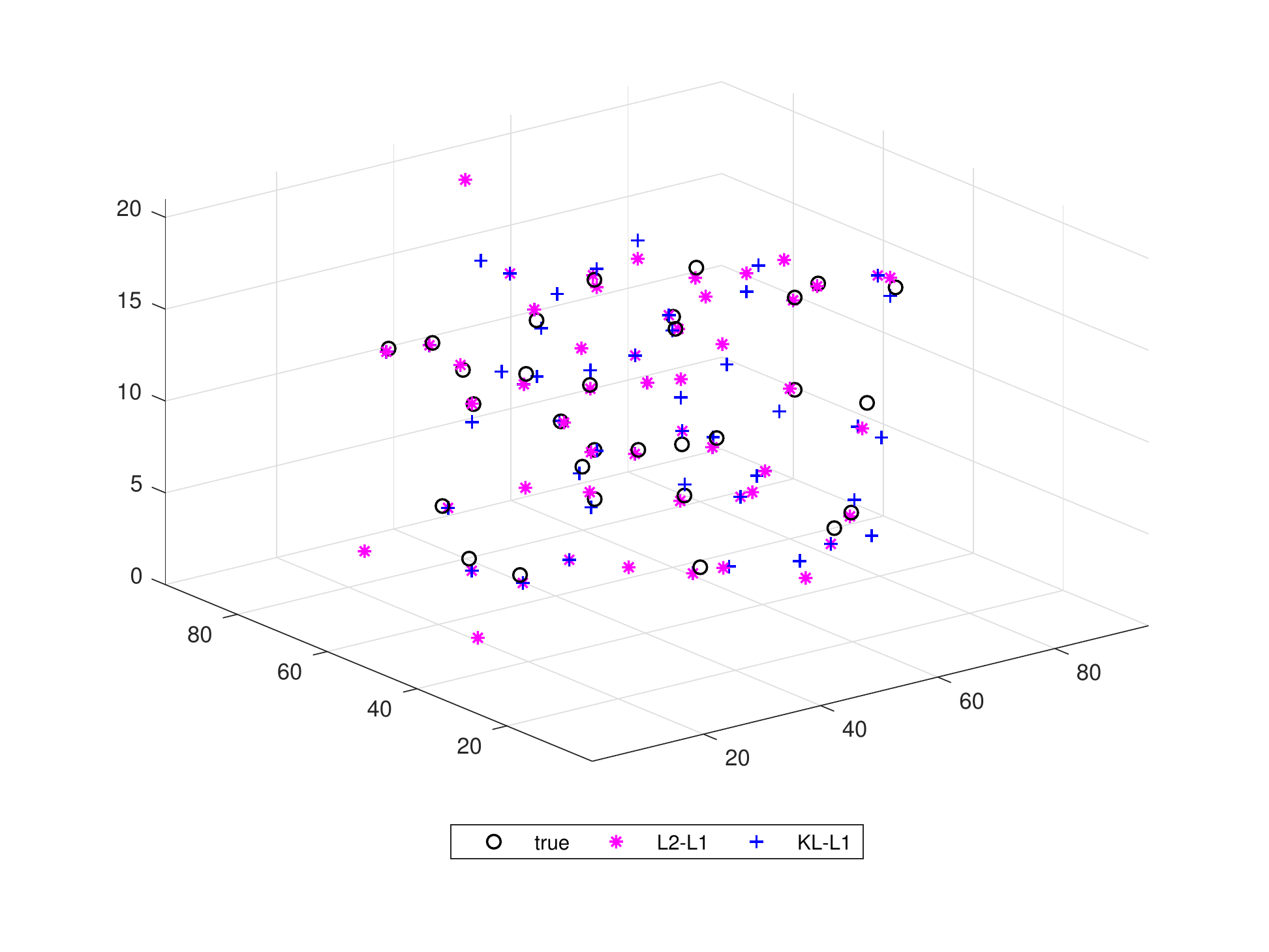}} 
\subfloat[With post-processing (non-convex regularization)]{\includegraphics[width=0.42\textwidth, height=0.25\textheight]{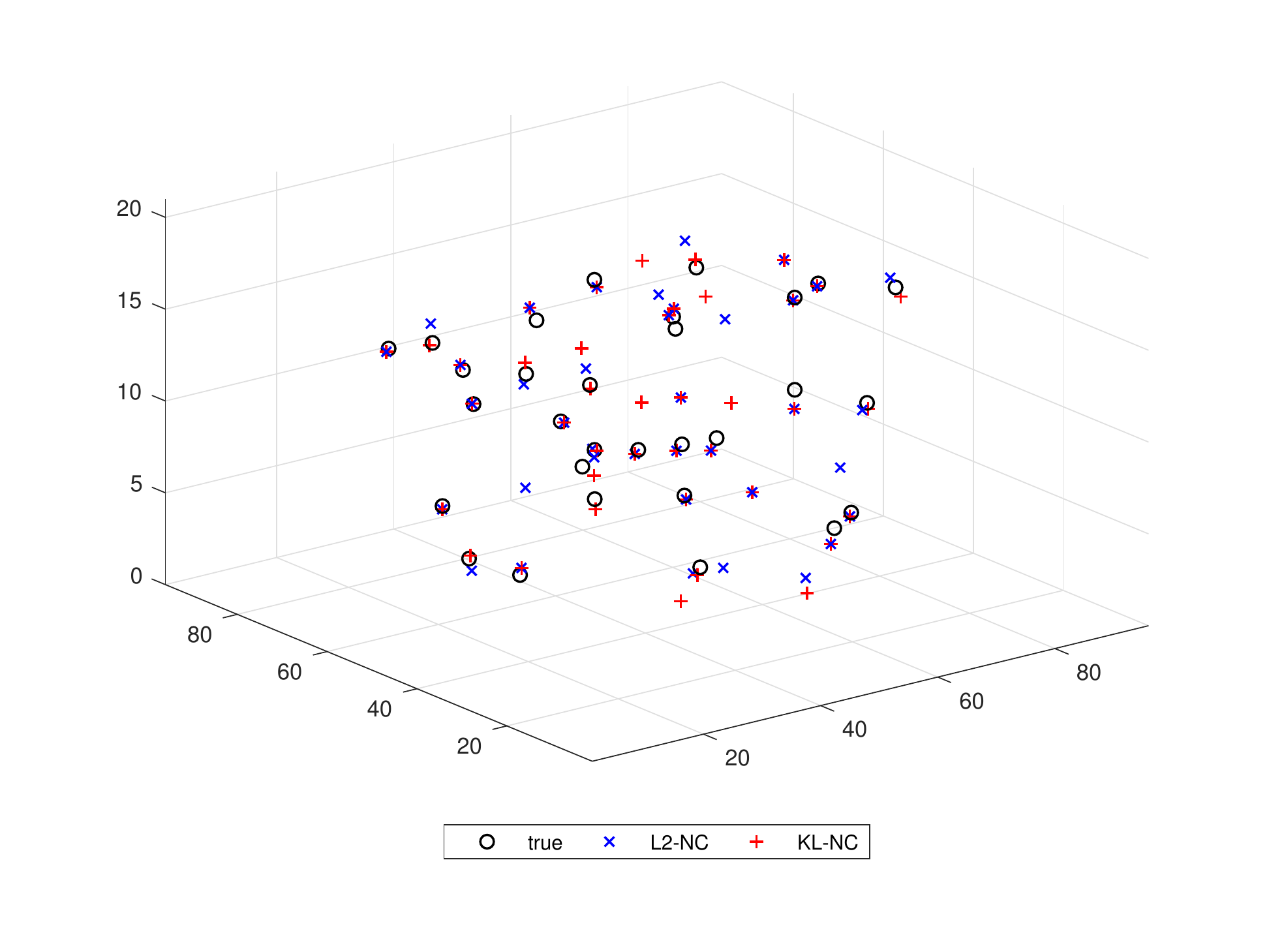}} 
\caption{Poisson noise case: 3D estimated results from 4 algorithms (30 point sources). } 
\label{fig:compare_distr_30}
\end{figure}

We  tested four algorithms for the Poisson noise case with a number of point source densities, namely 5, 10, 15, 20, 30 and 40, and computed the average recall and precision rates of 50 
images for each density and for each algorithm; see \Cref{Tab: with_postprocessing}. The results show  superior results  of our method in terms of both recall and precision rates, with the best recall and precision rates in each case  labeled by bold fonts.   As in the above discussion, our non-convex regularization tends to eliminate more false positives, and this increases the  precision rate. The KL data-fitting term, on the other hand, improves the recall rate as we see by comparing the results of KL-NC with $\ell_2$-NC. Before post-processing, we see that all the algorithms have low precision rates, especially  the two employing the $\ell_1$ regularization model at less than 10\%.

\begin{table}[htbp]
\begin{center}
\caption{Poisson noise case: Comparisons of  $\ell_2$-$\ell_1$,  $\ell_2$-NC and  KL-$\ell_1$ with our KL-NC.  }
\begin{tabular}{c|cc|cc|cc|cc}
\hline  &  \multicolumn{2}{c}{$\ell_2$-$\ell_1$} &  \multicolumn{2}{c}{$\ell_2$-NC}    &\multicolumn{2}{c}{KL-$\ell_1$} & \multicolumn{2}{c}{KL-NC}\\
\hline No.~Sources & Recall & Prec. & Recall & Prec. & Recall & Prec.  & Recall & Prec.    \\
\hline  5 & {\bf 100.00}\% & 68.91\%  & 97.60\% &  89.15\% & 98.93\% & 	58.64\% & {\bf 100.00}\% & {\bf  97.52}\%  \\
\hline 10 & {\bf 99.60}\% &  55.95\%   & 94.80\% &  83.51\% & 99.40\% &  65.24\% & { 99.40}\% & {\bf  93.69}\% \\
\hline 15 & {98.67}\% &  56.28\%  & 92.80\% &  84.77\% & {\bf 98.93}\% &  58.64\% & 98.40\% & {\bf  88.60}\%  \\
\hline 20 & 97.70\% &  56.50\%  & 95.20\% &  80.92\%   & {\bf 98.10}\% &  57.82\% & {97.70}\% & {\bf  87.49}\% \\
\hline 30  & {96.00}\% &  55.74\%  & 93.93\% &  77.77\% & {94.00}\% &  56.22\% & {\bf 96.20}\% & {\bf 79.75}\% \\
\hline 40  & 93.80\% &  52.68\%  & {\bf 95.40}\% &  59.34\% & 93.70\% &  54.29\% & {95.00}\% & {\bf 73.35}\% \\
\hline
\end{tabular}\label{Tab: with_postprocessing}
\medskip
\end{center}
\end{table}

\begin{figure}[htbp]
\centering
\subfloat[$\ell_2$-$\ell_1$]{\includegraphics[width=0.39\textwidth]{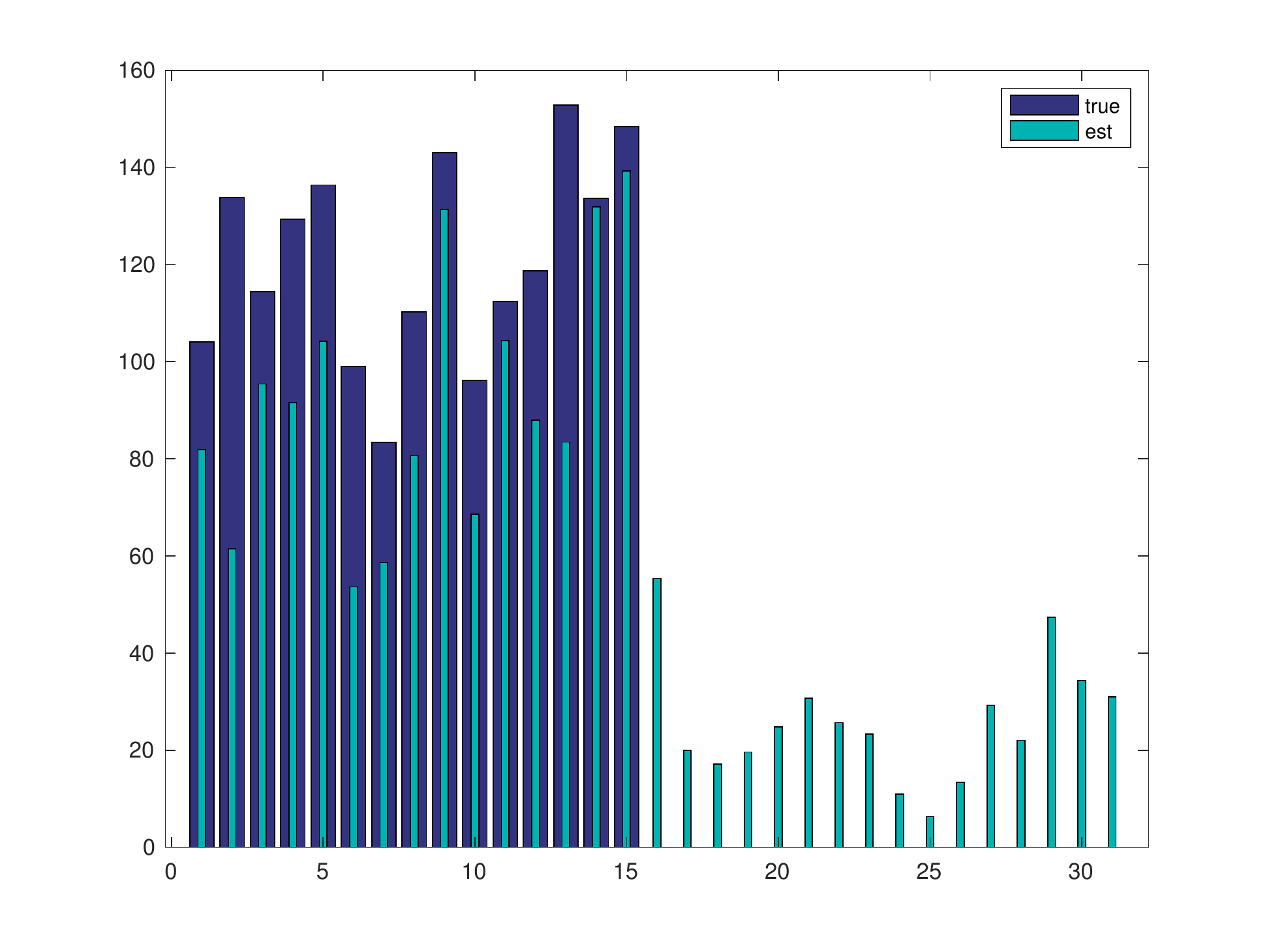}}
\hspace{0.01mm}
\subfloat[$\ell_2$-NC]{\includegraphics[width=0.39\textwidth]{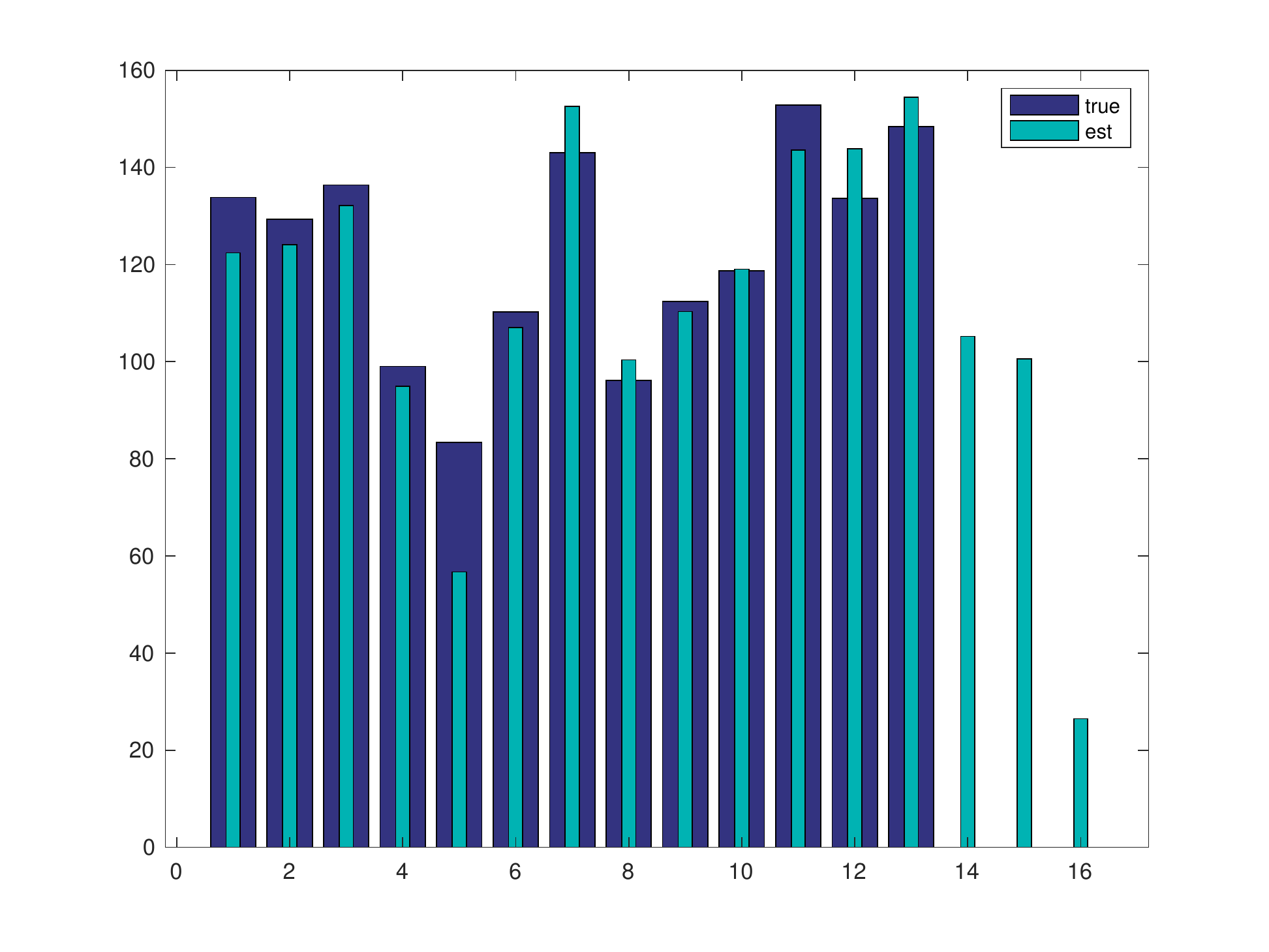}}
\hspace{0.01mm}
\subfloat[KL-$\ell_1$]{\includegraphics[width=0.39\textwidth]{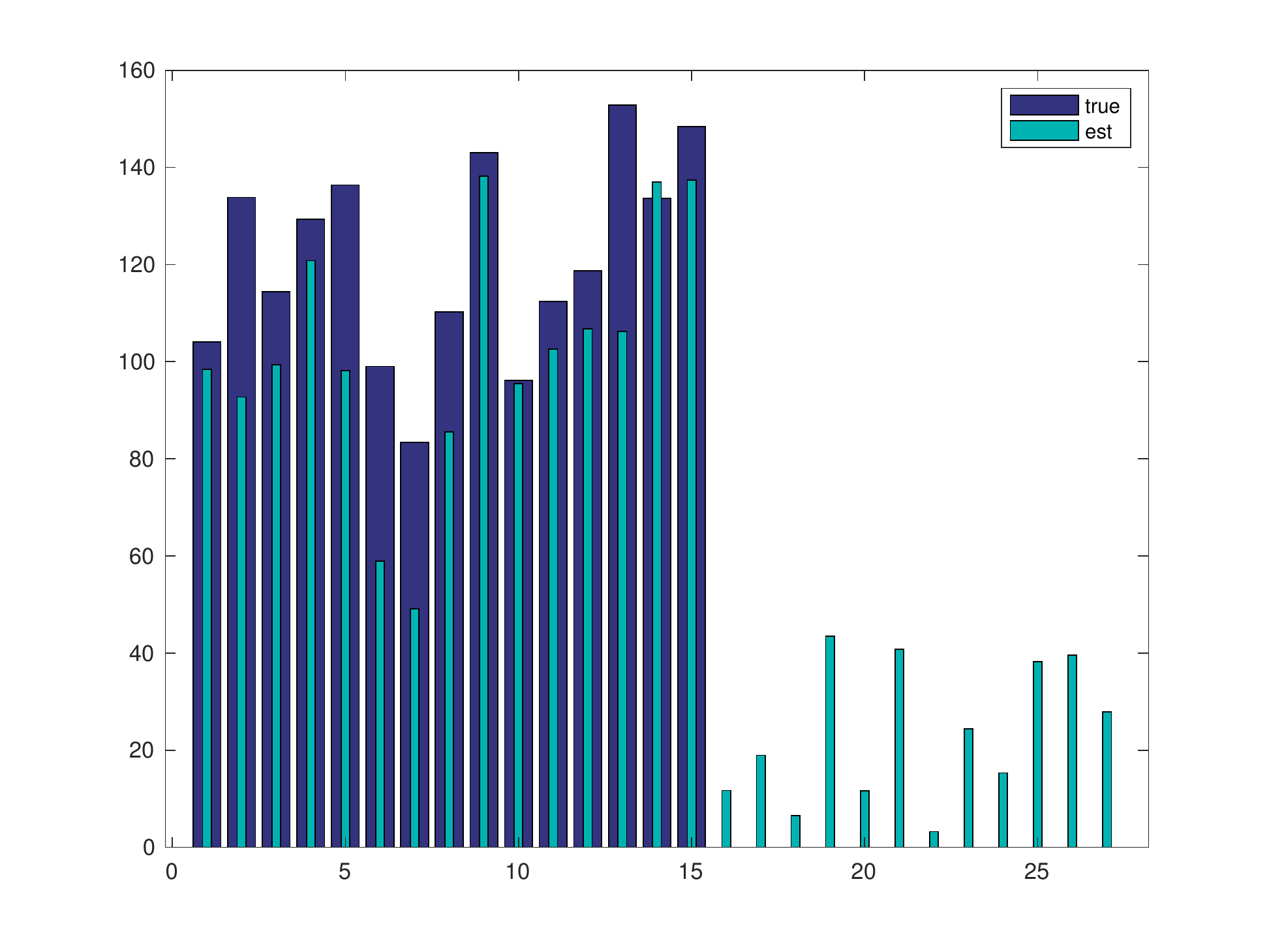}} 
\hspace{0.01mm}
\subfloat[KL-NC]{\includegraphics[width=0.39\textwidth]{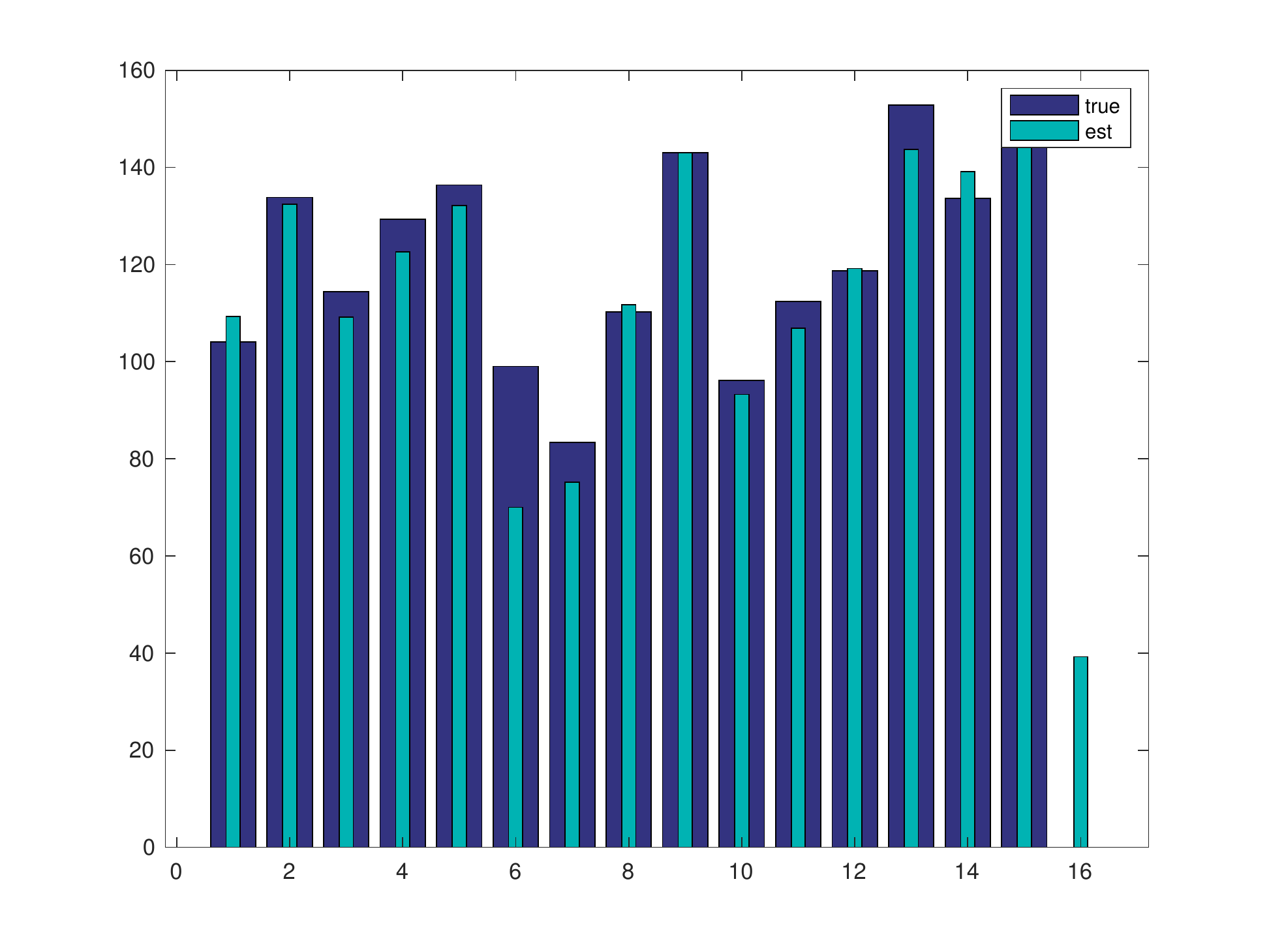}}
\caption{Poisson noise case:  Tests on estimating flux values. The bar graph with no value  ground truth part corresponds to a false positive. }\label{fig:flux}
\end{figure}

		We now compare the results of the estimations of the flux $f$  by these four algorithms, considering specifically the case of 15 point sources. In \Cref{fig:flux}, we plot the fluxes of ground truth as well as the fluxes of the estimated point sources for the true positive point sources. For the false positive point sources, we only  show 
the estimated fluxes.  Both $\ell_1$ models underestimate the fluxes. The rotating PSF images for false positives carry the energy away from the true positive source fluxes. In non-convex models, we also have similar observations when we have false positives. For example, in \Cref{fig:flux}(d), we see the flux on the fifth bar is  underestimated more than the others. We note that its rotating PSF is overlapping with the image of a false positive. 
The more false positives an algorithm recovers the more they will spread out the intensity, leading to more underestimated fluxes for the true positives. 

We also tested 50 different observed images for each density, and analyzed the relative error in the estimated flux values, which we define as 
\begin{equation*}
\mathrm{error} = \frac{\mathbf{f}_{\mathrm{est}}-\mathbf{f}_{\mathrm{tru}}}{\mathbf{f}_{\mathrm{tru}}},
\end{equation*}
where $(\mathbf{f}_{\mathrm{est}}, \ \mathbf{f}_{\mathrm{tru}})$ is the pair which contains the flux of an identified true positive and the corresponsing  ground truth flux. In \Cref{fig:30flux_per}, we plot the histogram of the relative errors on these four optimization models in the 30 point sources case. We still see the advantage of KL-NC over other algorithms in this respect. The distribution of relative errors mostly lies within $[0, \ 0.1]$. For the $\ell_1$ regularization algorithms, the distribution of the relative error spreads out and there are  many cases with error higher than 0.3.  


\begin{figure}[t!]
\centering
\subfloat[$\ell_2$-$\ell_1$]{\includegraphics[width=0.39\textwidth]{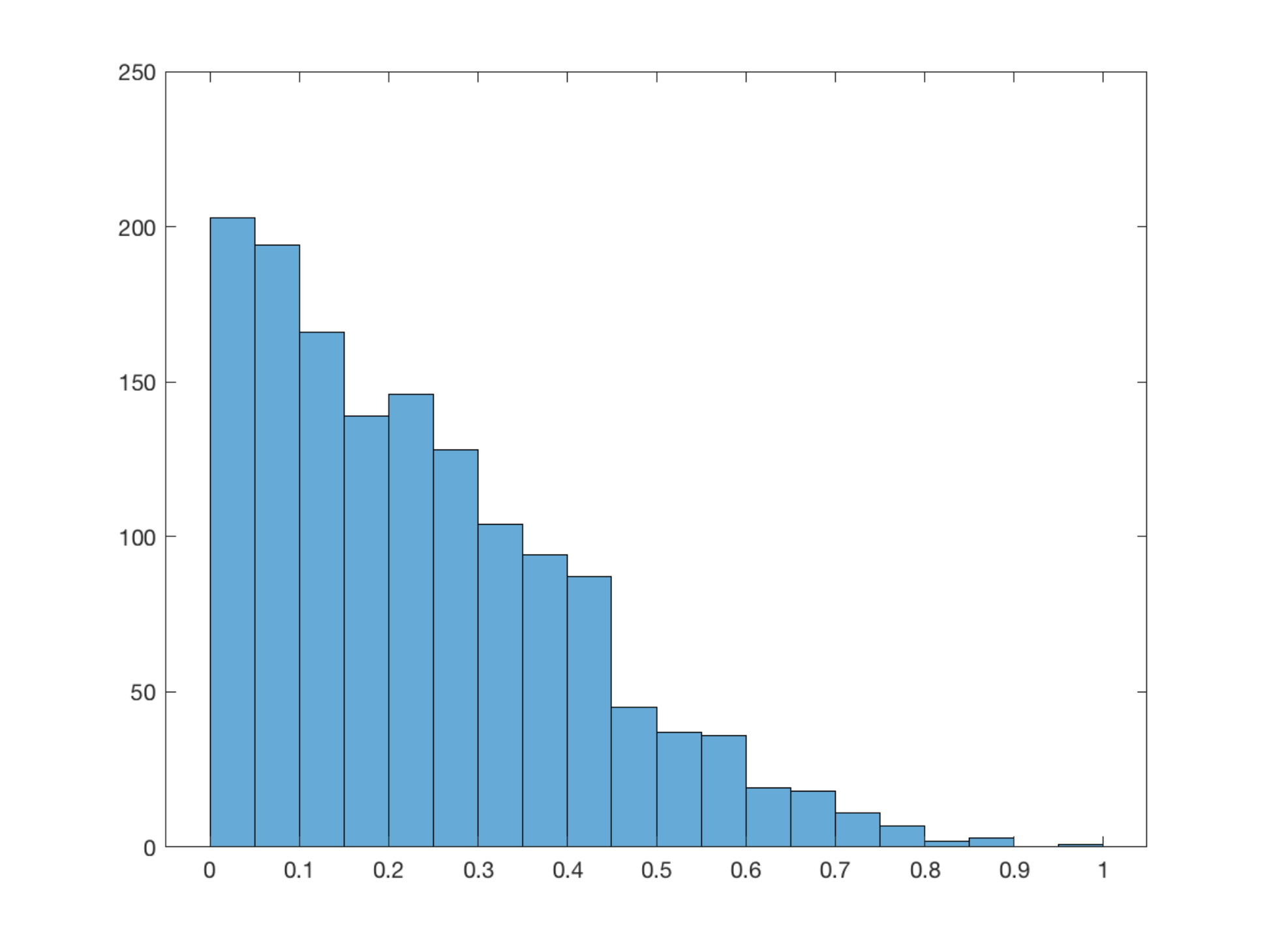}}
\hspace{0.01mm}
\subfloat[$\ell_2$-NC]{\includegraphics[width=0.39\textwidth]{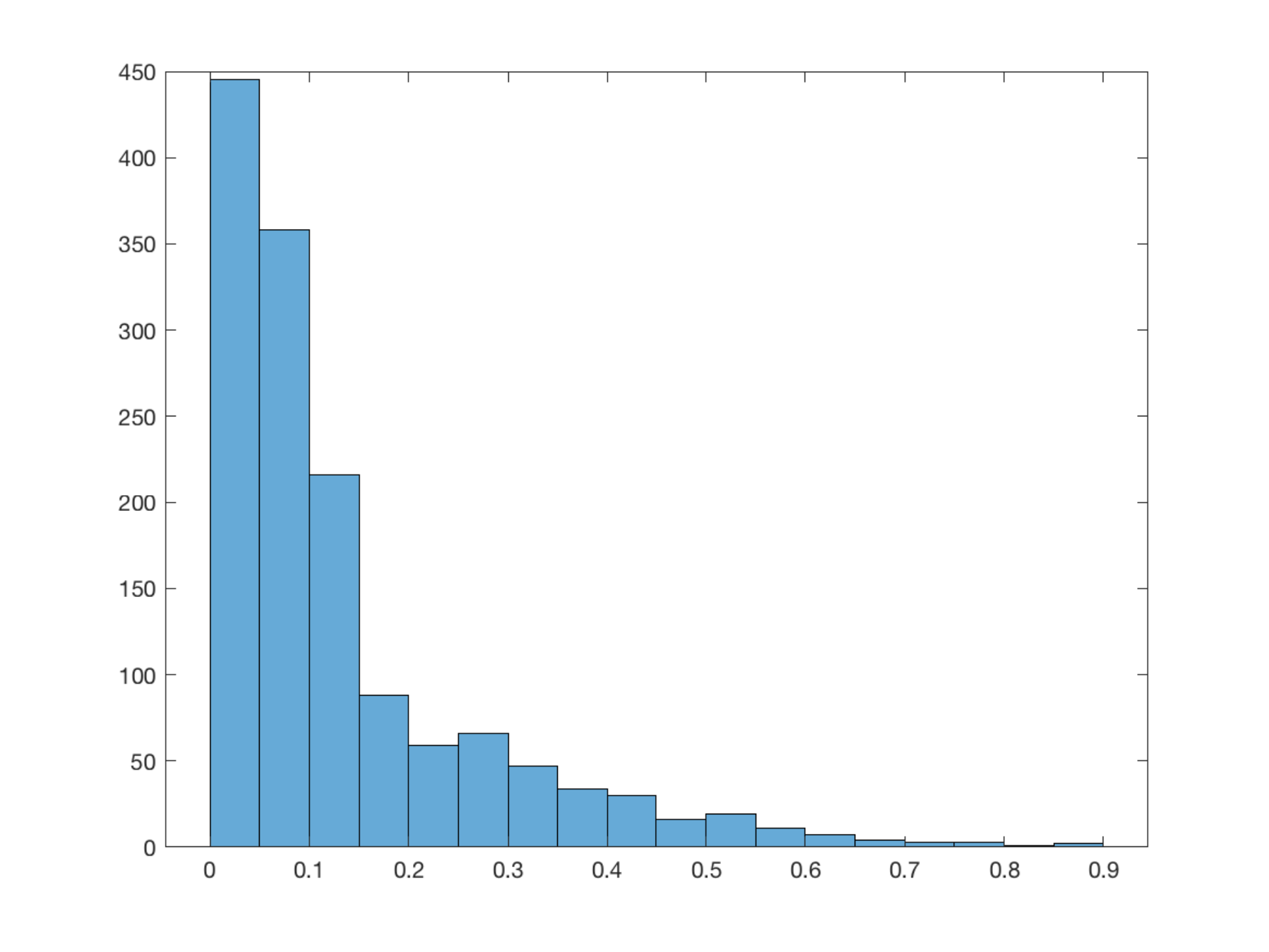}}
\hspace{0.01mm}
\subfloat[KL-$\ell_1$]{\includegraphics[width=0.39\textwidth]{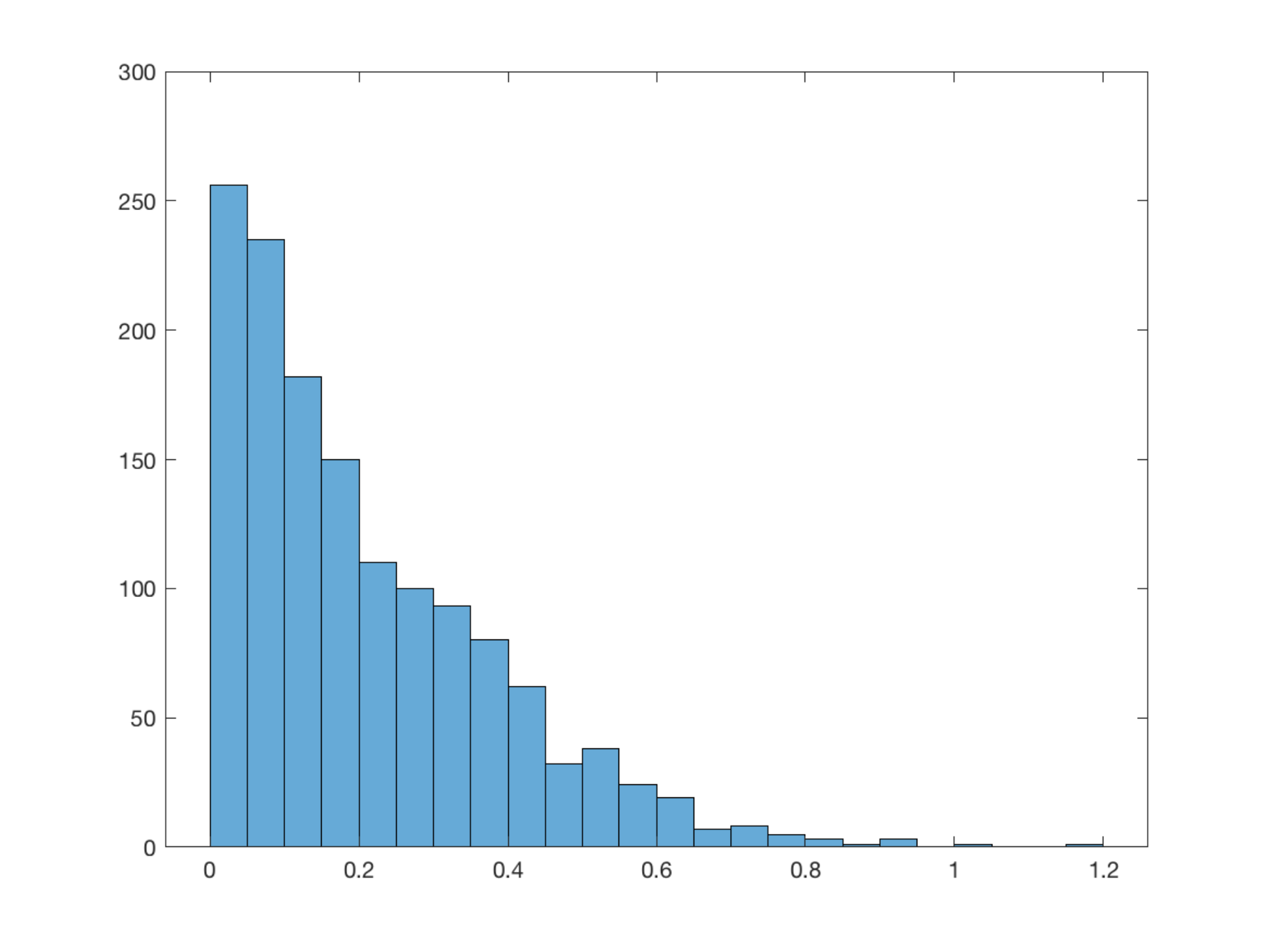}} 
\hspace{0.01mm}
\subfloat[KL-NC]{\includegraphics[width=0.39\textwidth]{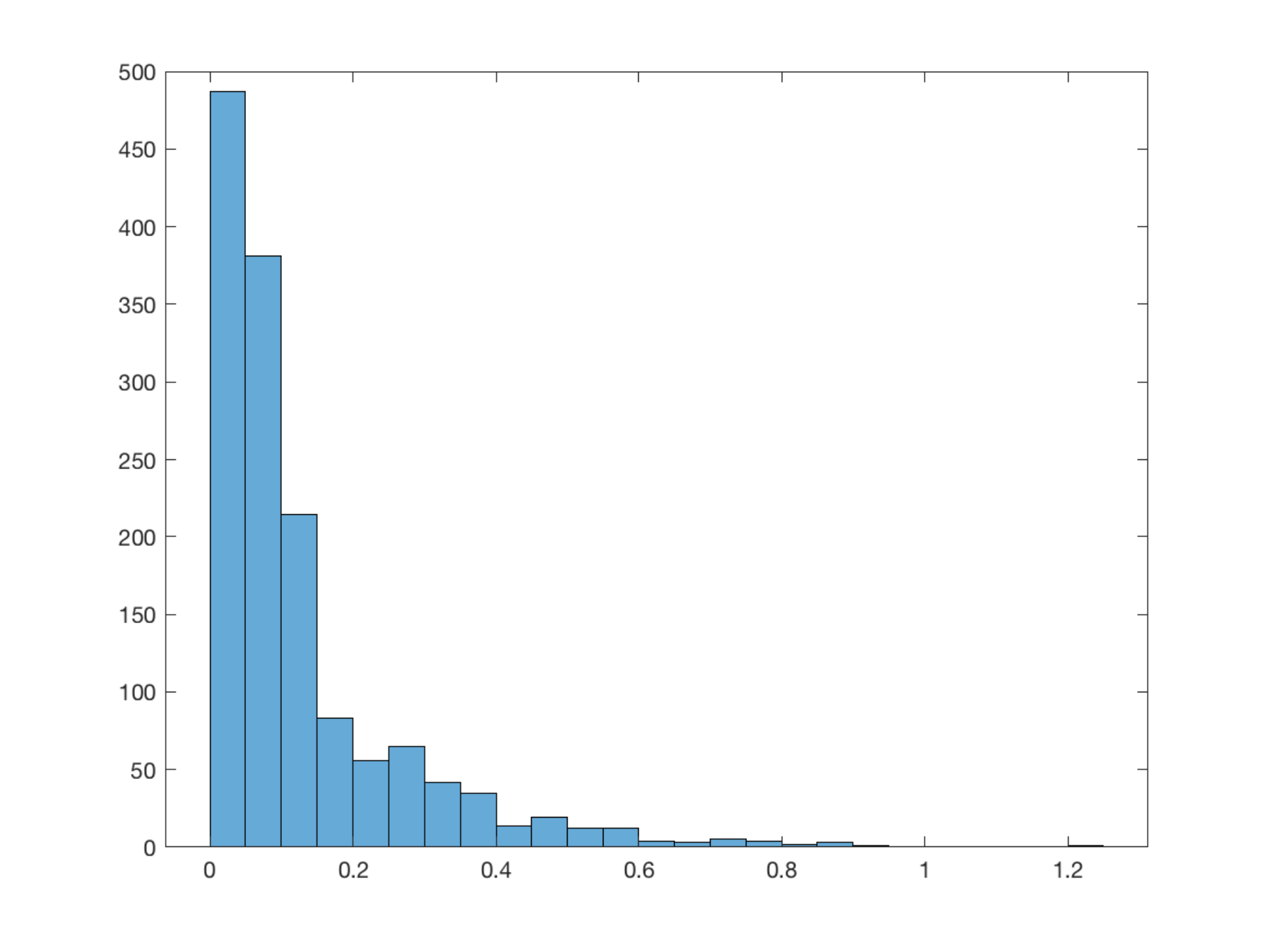}}
\caption{Histogram of relative errors of flux values in the 30 point sources case. } 
\label{fig:30flux_per}
\end{figure}

\section{Conclusions and future work}\label{sec:Conclusions}
We have proposed  non-convex optimization algorithms for the 3D localization of a swarm of randomly spaced point sources using a rotating PSF which has a single lobe in the image of each point source. It has advantages over the double-lobe rotating PSF, e.g.  \cite{Rice2016generalized,moerner2015single,DH2008pavani,DH2009pavani}, especially in cases where the point source density is high.
 In addition, for the Poisson-noise case we have proposed a new iterative scheme for refining the estimates of the source fluxes after the sources have been localized.

These techniques can be applied to other rotating PSFs as well as other depth-encoding PSFs for accurate 3D localization and flux recovery of point sources in a scene from its image data under {both the Gaussian and Poisson noise models}. 
Applications include not only 3D localization of  space debris, but also super-resolution 3D single-molecule localization microscopy, e.g.  \cite{Book_SR_micro2017,3dsml2017review}.
 Tests of our algorithms based on real data collected using phase masks fabricated for both applications are currently being planned.  In addition,  work involving
snapshot multi-spectral imaging, which will permit accurate material characterization, as well as higher 3D
resolution and localization of space micro-debris via a sequence of snapshots, is underway.

\acknowledgments     
The authors acknowledge funding support for the work from the US Air Force Office of Scientific Research under grant FA9550-15-1-0286, and from  Hong Kong grants (HKRGC Grant CUHK14306316, HKRGC CRF Grant C1007-15G, HKRGC AoE Grant AoE/M-05/12, CUHK DAG 4053211, and CUHK FIS Grant  1907303). 


\bibliography{ref_rPSF}   
\bibliographystyle{ieeetr}   

\end{document}